\documentclass[journal]{IEEEtran}
\pdfoutput=1
\usepackage{graphicx}
\usepackage{amsfonts}
\usepackage{amssymb}
\usepackage[cmex10]{amsmath}
\begin{document}
\title{Zero-Crossing Waveform 
	Interferometry\\ 
	{\huge An Alternative to Correlation in Signal Processing}
}
\author{W.~J.~Szajnowski~\IEEEmembership{}
        {}
\thanks{W. J. Szajnowski is a Visiting Professor at 
	Centre for Vision, Speech and Signal Processing, University of Surrey, Guildford GU2 7XD, U.K., e-mail: w.j.szajnowski@surrey.ac.uk}
\thanks{}
}

\markboth{ }%
{Shell \MakeLowercase{\textit{et al.}}: Bare Demo of IEEEtran.cls for Journals}

\maketitle

\begin{abstract}
\boldmath
 It is shown that multiple representations (such as replicas or Hilbert transforms) of a random waveform can interfere constructively to form a compact pattern, akin to a wave packet, when the representations are created in synchrony with zero crossings of the waveform. A function of such 'engineered' zero-crossing interferograms can exhibit time-delay resolution superior to that associated with a conventional correlation function, especially for waveforms with slowly-decaying power spectra. A phenomenon of local slew rate at zero crossings is exploited to substantially reduce the Cram\'er-Rao bound on time-delay estimators. A system, based on a
 concept of elapsed time, is proposed to determine zero-crossing interferograms in real time.
 \end{abstract}


\IEEEpeerreviewmaketitle

\section{Introduction}
\IEEEPARstart{I}{n} science and engineering, the mathematical concept of correlation between a pair of observables is exploited to investigate a statistical or functional relationship between these observables. As a consequence, the operation of various scientific instruments and methods is based on some form of correlation [1], [2]. However, correlation, being a mathematical construct, is not well suited to investigating a broad class of physical phenomena and 
processes as they are evolving in real time.

One important area in which correlation-based techniques play a prominent role is that of localisation and imaging of non-cooperative objects in some specified region of interest. Such tasks can be performed by an active or passive system incorporating sensors which can extract useful information by collaborative processing signals reflected by the objects. In an active system, the surveillance region is illuminated by an interrogating energy waveform generated by the system, whereas in passive sensing, illumination energy is provided by uncontrolled sources of ambient noise of seismic, acoustic or electromagnetic origin [3]--[7]. 

In practice, intercepted signals, in addition to occupying a wide frequency band, may also manifest a non-stationary and chaotic nature with identifiable intermittent transients. Furthermore, when a physical phenomenon is to be processed in real time, the conventional notion of {\emph {chronological}} time, extending from minus infinity to plus infinity, has to be replaced by a semi-infinite measurable quantity of {\emph {elapsed}} time that starts at a continually changing moment of a present event and 'moves backward' to include all already observed past events [8], [9]. As a consequence, many standard  correlation techniques based, explicitly or implicitly, on the assumptions of stationarity and conventional notion of time can only be of limited practical use. 

Despite the many advances in correlation-based signal processing, the performance of any current man-made sensing system is vastly inferior to that exhibited by echolocating mammals and birds. 
The recognition of this performance gap between man-made and biological systems has triggered a renewed interest in studying physical and biological principles of signal representation and processing [10]--[15].

\subsection{Preliminary Considerations}
Assume that $x(t)$ is a reference signal waveform, such as that representing random noise, and let $x(t+\tau)$ be its replica shifted in time by $\tau$. A measure of dissimilarity between these two waveforms can be defined by
 \begin{IEEEeqnarray}{lll}
	\widehat{\mathsf{D}}(\tau)\,\,\,&&\triangleq\,
	\left \langle  [x(t)-x(t+\tau) ]^2 \right \rangle\\
	&&=\,\langle x^2(t) \rangle + \langle x^2(t+\tau) \rangle 
	-\,2\mspace{1mu}\langle x(t)\mspace{1mu}x(t+\tau) \rangle
\end{IEEEeqnarray}
where $\langle \mspace{1mu} \cdot \mspace{1mu} \rangle$ denotes the average over a time interval. The measure $\widehat{\mathsf{D}}(\tau)$ is then a second-order moment of the increment $[x(t)-x(t+\tau)]$. 

The above function, referred to as a second-order {\em {structure function}}, was first introduced by Kolmogorov in a study on locally isotropic and homogeneous turbulence [16]. Also, Woodward, in an analysis of time resolution in radar [17], proposed the same form (1) as a measure of departure of $x(t)$ from $x(t+\tau)$.

In the case of a stationary signal waveform, the first two terms in (2) are estimates of the same quantity, power of $x(t)$, independent of $\tau$. Therefore, when the mean of $x(t)$ is assumed to be zero, it has become a common practice to use an empirical correlation function, 
\begin{equation}
\widehat{R}_{XX}(\tau) \, \triangleq \,
	\langle x(t)\mspace{1mu}x(t+\tau) \rangle
\end{equation}
as a measure of association between a waveform $x(t)$ and its time-shifted replica $x(t+\tau)$. 

For a finite averaging interval, even when the mean of $x(t)$ is zero, a corresponding empirical mean (the average) will fluctuate from interval to interval; hence, the structure function may be a preferred choice [18]. However, in contrast to (1), an empirical correlation function (3) can be exploited to detect and resolve multiple time-shifted replicas, each having a different power and a different time shift.

Assume that a zero-mean reference signal waveform $x(t)$ has a substantial number of zero crossings occurring at non-uniform, possibly non-deterministic, time instants $\{t_i\}$. Then, a {\emph {local}} structure function  $\widehat{\mathsf{D}}_0(\tau)$, obtained from (2), can be defined by
\begin{equation}
\widehat{\mathsf{D}}_0(\tau) \, \triangleq \, {\left
\langle x^2(t_i+\tau) \right \rangle}_{\mspace{-2mu}{\tt {C}}}
\end{equation}
where ${\langle\mspace{1mu} \cdot\mspace{1mu} \rangle}_{\mspace{-2mu}{\tt {C}}}$ denotes the average over a set of indices, ${\tt {C}} = \{i\}$, of zero crossings occurring at $\{t_i\}$.

A value of $\widehat{\mathsf{D}}_0(\tau)$ at any specified time instant, say, $\tau=\tau_s$, is determined by sampling $x(t)$ at the times $\{t_i+\tau_s\}$ and averaging the corresponding values, $\{x^2(t_i+\tau_s)\}$, to produce $\widehat{\mathsf{D}}_0(\tau_s)$. Obviously, this sampling and averaging process may be repeated for a number of different values of $\tau_s$ to construct a discrete-time representation of $\widehat{\mathsf{D}}_0(\tau)$.

The local structure function $\widehat{\mathsf{D}}_0(\tau)$ can also be determined by {\em {shifting and aligning}} in time multiple copies $\{x_i(\tau)\}$ of $x(t)$, where $x_i(\tau) \triangleq x(t_i+\tau)$, so that  the zero-crossing instants $\{t_i\}$ will all have collapsed onto a single point $\tau=0$. Then, $\widehat{\mathsf{D}}_0(\tau)$ is obtained by averaging the functions $\{x^2_i(\tau)\}$ of relative time $\tau$.

In order to demonstrate an important difference between the global (1) and local (4) structure functions, assume that $x(t)$ is a realization of a real-valued stationary Gaussian random process $X(t)$ with zero mean, variance $\sigma^2_X$, and  correlation coefficient $r_X(\tau)=R_{XX}(\tau)/\sigma^2_X$, 
where $R_{XX}(\tau)$ is the correlation function corresponding to its empirical counterpart~(3). 

 As a consequence of the ergodic theory, when the averaging time interval tends to infinity{\footnote{For a finite zero-crossing rate, as the number of zero crossings in (4) is approaching infinity, so is the time interval including these crossings.
}}, 
the averages (1) and (4) will converge with probability one to corresponding ensemble means, $\mathsf{D}(\tau)$ and $\mathsf{D}_0(\tau)$. In the considered case,
 $$
 \mathsf{D}(\tau)\, = \,
 2\mspace{1mu} \sigma^2_X\mspace{-1mu} \left [1 -r_X(\tau) \right ] 
 $$
 whereas the mean $\mathsf{D}_0(\tau)$, determined from a second moment of the Slepian process (Section II.\,B), can be expressed as
 $$
 \mathsf{D}_0(\tau)\,=\,\sigma^2_X\mspace{-2mu} \left [
 1 - r^2_X(\tau) - |r^{\prime}_X(\tau)|^2/r^{\prime\prime}_X(0)
  \right ].
 $$
 \begin{figure}[]
 	\centering
  	\includegraphics[width=6.8cm]{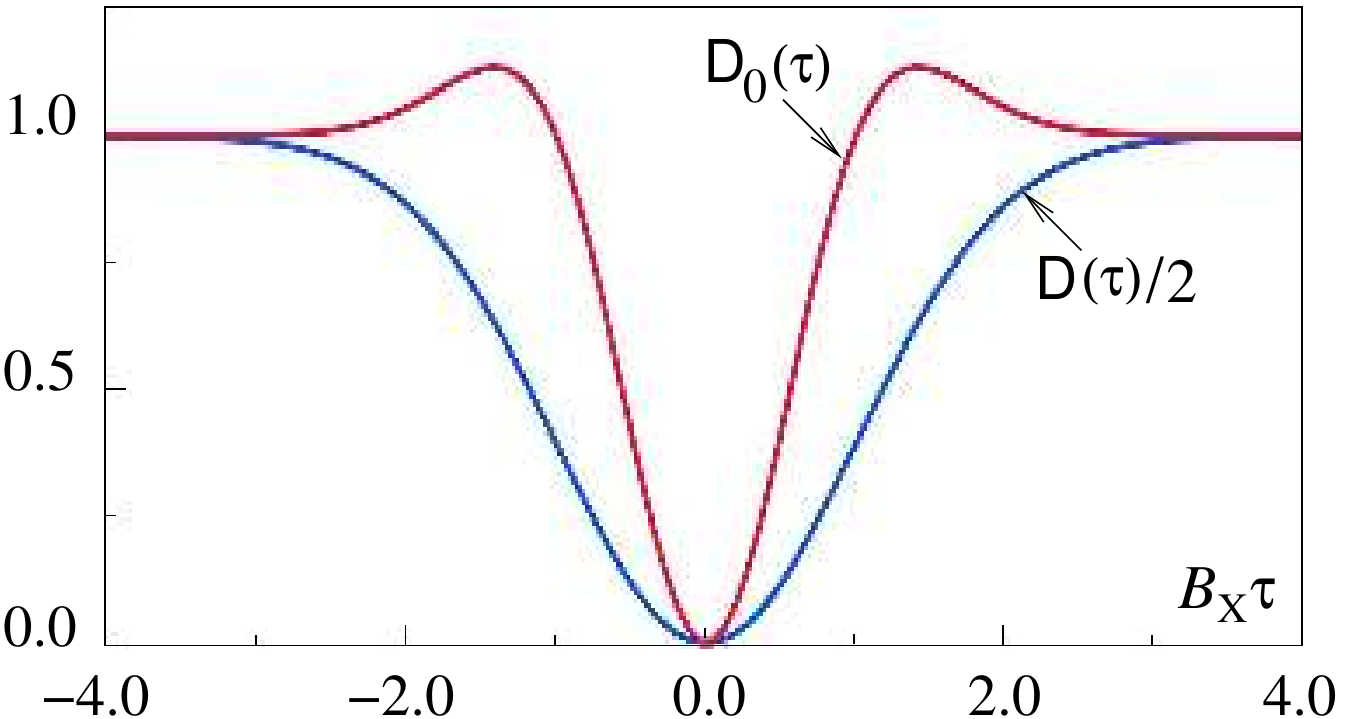}
 	\vspace{-0.2cm}
 	\caption{Examples of local, $\mathsf{D}_0(\tau)$, and global, $\mathsf{D}(\tau)/2$, structure functions.}
 \end{figure}
For illustrative purposes, Fig.\,1 shows the local, $\mathsf{D}_0(\tau)$, and adjusted global, $\mathsf{D}(\tau)/2$, structure functions, when $\sigma^2_X\!=\!1$, and the correlation coefficient of $X(t)$ has a Gaussian shape, $r_X(\tau) = \exp(-B_X^2\mspace{1mu}\tau^2/2)$, where 
$B^2_X = -r^{ \prime \prime}_X(0)$ is the mean-square (ms) bandwidth of $X(t)$. As seen, the local function (4) may be preferred to the adjusted global function $\mathsf{D}(\tau)/2$.

A set ${\tt {C}}$ of indices of zero crossings, appearing in the local structure function (4), is a sum of two subsets: ${\tt {C}}^+$, containing indices $\{k\}$ of zero {\emph {upcrossings}} at times $\{t_k\}$, and ${\tt {C}}^-$, containing indices $\{\ell\}$ of zero {\emph {downcrossings}} at times $\{t_{\ell}\}$. 

Local increments, $\{x(t_k+\tau)\}$, associated with  
zero upcrossings, can be used to determine the average
\begin{equation}
{\left	\langle x(t_k+\tau) \right \rangle}_{\mspace{-2mu}{\tt {C}}^+}\,.
\end{equation}
In a similar manner, the average 
\begin{equation}
{\left	\langle x(t_{\ell}+\tau) \right \rangle}_{\mspace{-2mu}{\tt {C}}^-}
\end{equation}
can be formed by using increments, $\{x(t_{\ell}+\tau)\}$, associated with zero downcrossings. The averages, (5) and (6), can be regarded as  empirical {\em {waveform interferograms}}, associated, respectively, with zero  upcrossings and zero downcrossings.

\subsection{Objectives}
The main objective of this paper is to analyse zero-crossing waveform interferograms, (5) and (6), and use them as the building blocks of  two new functions that can replace an empirical correlation function (3) in applications related to object localisation and imaging.

It is a further objective of this paper to determine a resolution gain, when the proposed functions are used instead of the conventional correlation function (3).

\section{Zero-Crossing Waveform Interferograms}
Let $X(t)$ be a real-valued, stationary and ergodic random process with zero mean, $\mathrm{E}\{X(t)\}=0$, and variance $\mathrm{E}\{X^2(t)\} = \sigma^2_X$, where $\mathrm{E}\{\cdot\}$ denotes statistical expectation. 
It is assumed that realizations, 
 $\{x(t)\}$, of the process $X(t)$ are continuous and differentiable, with distinct and reliably detectable zero crossings. The process is also assumed to have a finite number of zero crossings in any finite time interval. 
 
A random process can be regarded as an ensemble of an infinite number of realizations. Realizations of a random process $X(t)$ will interfere {\emph {constructively}} within a specified time interval, when the realizations are suitably aligned (or synchronised) in time; otherwise, their superposition will produce a constant of zero value. 
 
 Such temporal alignment of realizations can be accomplished by defining a {\em {significant event}}, such as a zero crossing, that may only occur at some distinct and detectable time instant. Then, a superposition of  realizations, {\emph {conditioned}} on that event, may produce a non-zero pattern, appearing in a time interval that includes the significant event.

A pattern resulting from the interference of such time-aligned realizations of $X(t)$ will be equal to the {\em {conditional mean}}  $\mathrm{E}\{X(t)|\mspace{1mu}\mathcal{B}\}$,
where $\mathcal{B}$ denotes a significant event of zero crossing (either upcrossing or downcrossing). More specifically, the underlying significant events are defined as follows:\\
for an upcrossing at a time instant $t_k$,
\begin{equation}
\mathcal{B}^+ \mspace{1mu}\triangleq\, 
\{X(t_k) = 0, \quad {\mathrm{and}} \quad  X'(t_k) > 0   \}
\end{equation}
and for a downcrossing at $t_{\ell}$,
\begin{equation}
\mathcal{B}^- \mspace{1mu} \triangleq\, 
\{X(t_{\ell}) = 0, \quad {\mathrm{and}} \quad  X'(t_{\ell}) < 0   \}.
\end{equation}
Accordingly, each of the two corresponding conditional means, $\mathrm{E}\{X(t)|\mspace{1mu}\mathcal{B}^+\}$ and $\mathrm{E}\{X(t)|\mspace{1mu}\mathcal{B}^-\}$, will be associated with a respective type of zero crossings.

From the ensemble point of view, conditioning on a zero crossing has no meaning, since the event $\mathcal{B}$, such as $\mathcal{B}^+$, is of probability zero. One way to define this kind of conditioning is to consider a family of events 
$\{\mathcal{B}^+_{\epsilon}\}$ with $\Pr(\mathcal{B}^+_{\epsilon})>0$, such that 
$\bigcap_{\mspace{2mu}\epsilon>0}\mathcal{B}^+_{\epsilon} = \{X(t_k) = 0,X'(t_k) > 0\}$. Then, the conditional probability 
$\Pr(\cdot\mspace{1mu}|\mspace{1mu}\mathcal{B}^+)$ is defined as the limit 
$\lim_{\mspace{1mu}\epsilon \rightarrow  0}\Pr(\cdot\mspace{1mu}|\mspace{1mu}\mathcal{B}^+_{\epsilon})$. Unfortunately, this limit will depend on the construction of the family 
$\{\mathcal{B}^+_{\epsilon}\}$. This result, known as the Kac-Slepian paradox, shows that the classical structure of probability space is insufficient to define such a conditioning, and the physics of the problem under consideration must be taken into account [19], [20]. 

As discussed in more detail in [19], the definition of the event $\{\mathcal{B}^+_{\epsilon}\}$ has a physical meaning, when it is assumed 
that a random process $X(t)$ has a zero upcrossing somewhere in the {\em {horizontal window}} $(t_k - \epsilon/2 < t < t_k + \epsilon/2)$, as $\epsilon \rightarrow 0$; the event 
$\{\mathcal{B}^-_{\epsilon}\}$ of a zero downcrossing can be defined in a similar manner.

A different approach to dealing with level-crossing conditioning is presented in [20]. The proposed method is an adaptation of the theory of pull-back of distributions in stochastic calculus.

\subsection{Crosslation ${\mspace{-4mu}}^2$ Function}
In the physical world, only a single realization, a waveform $x(t)$, of the underlying random process $X(t)$ is available for processing.  
 Since the process $X(t)$ is assumed to be stationary and ergodic, a suitably defined average, determined from a single waveform $x(t)$, will converge to the corresponding (ensemble) conditional mean, such as $\mathrm{E}\{X(t)|\mspace{1mu}\mathcal{B}^+\}$, with probability one as the number of zero crossings approaches infinity. 

Assume that $\{t_k\}$ are the time instants of zero upcrossings by $x(t)$ and attach to each $t_k$ a {\emph {local}} increment function, defined by $x^+_k(\tau) \triangleq x(t_k+\tau)$, and referred to as a {\emph {crossjectory\,}}{\footnote {
The terms, {\emph {crosslation}} and {\emph {crosslator}}, coined by the author, are trade marks of Mitsubishi Electric Information Technology Centre Europe B.V.; the term {\emph {crossjectory}} is a neologism.}}. Hence, each crossjectory is simply a time-shifted copy of the entire waveform $x(t)$. By construction, the time instants $\{t_k\}$ will all have collapsed onto a single point $\tau = 0$, and the corresponding crossjectories $\{x_k^+(\tau)\}$ will share the same origin of relative time $\tau$.

When $k$ is running through the set of zero upcrossings, a waveform $x(t)$  will generate a sequence $\{x_k^+(\tau),\,\, k=1,2,\ldots\}$ of crossjectories. Each crossjectory $x_k^+(\tau)$ can be regarded as a realization of a conditional process $X^+(\tau) \triangleq X(t)|\mspace{1mu}\mathcal{B}^+$, where the conditioning event $\mathcal{B}^+$ is defined by (7). 

\begin{figure}
	\centering
	\includegraphics[width=6.8cm]{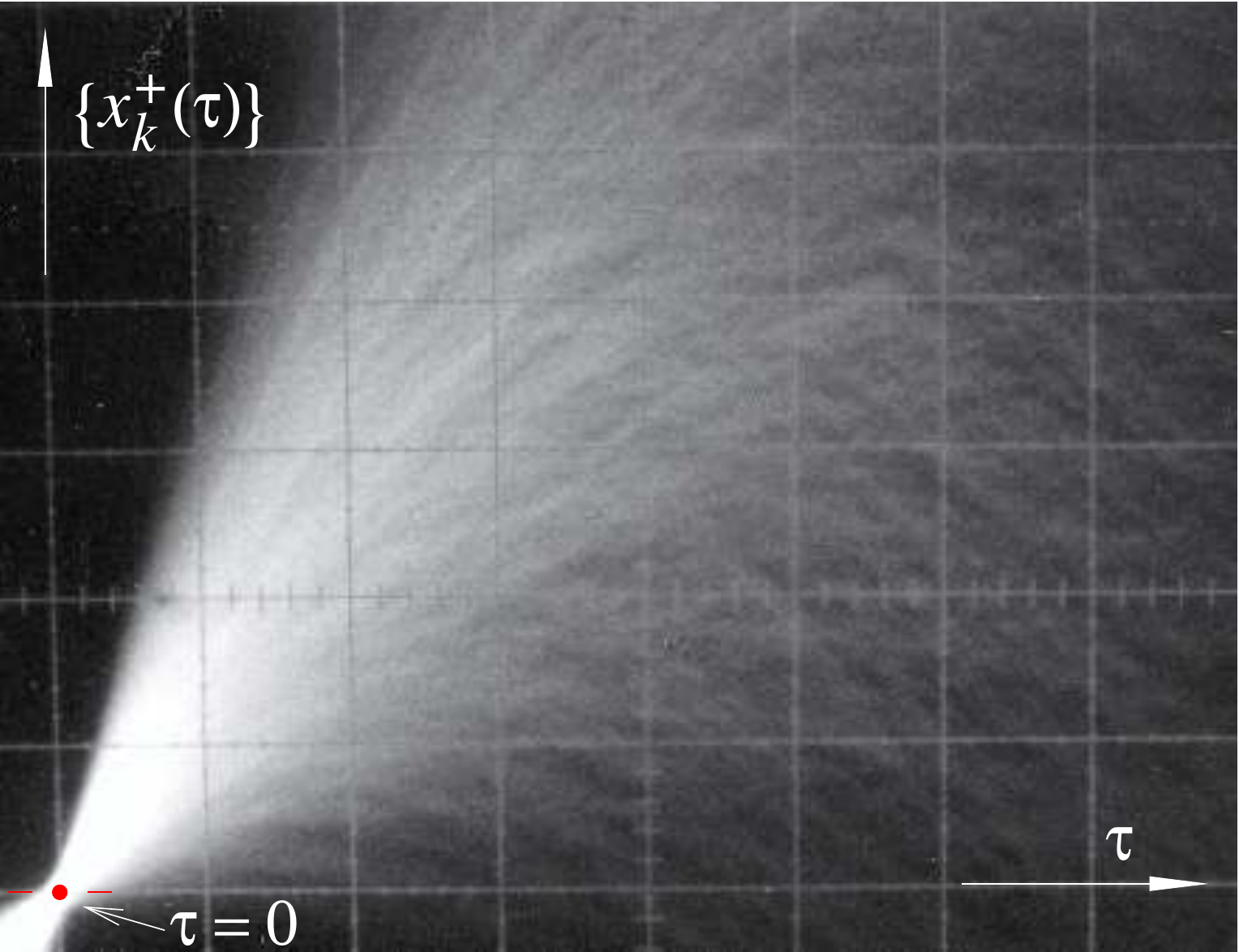}
	\vspace{-0.2cm}
	\caption{Crossjectories $\{x^+_k(\tau)\}$ of low-pass noise for $\tau \ge 0$.}
\end{figure}

Fig.\,2 is a long-exposure photograph of an analogue oscilloscope screen showing segments of noise crossjectories associated with zero upcrossings. The oscilloscope was driven by zero-mean low-pass noise, and triggered by its zero upcrossings. As seen, in the neighbourhood of $\tau = 0$, predominantly positive noise crossjectories will give rise to a non-zero average, being an estimate of the conditional mean, $\mathrm{E}\{X(t)|\mspace{1mu}\mathcal{B}^+\}$, for $\tau \ge 0$. 

Consider the following average,
\begin{equation}
\widehat{C}^+(\tau) \, \triangleq \,
 \frac {1}{n^+} \sum_{k=1}^{n^+} \mspace{-1mu}x_k^+(\tau) 
 \end{equation}
where $n^+$ is the number of upcrossings occurring in a time interval $T$. The average $\widehat{C}^+(\tau)$, viewed as an estimate of the ensemble mean ${C}^+(\tau) \equiv \mathrm{E}\{X^+(\tau)\}$, is then an empirical zero-upcrossing waveform interferogram (5).

\begin{figure}[t]
	\centering
	\includegraphics[width=6.9cm]{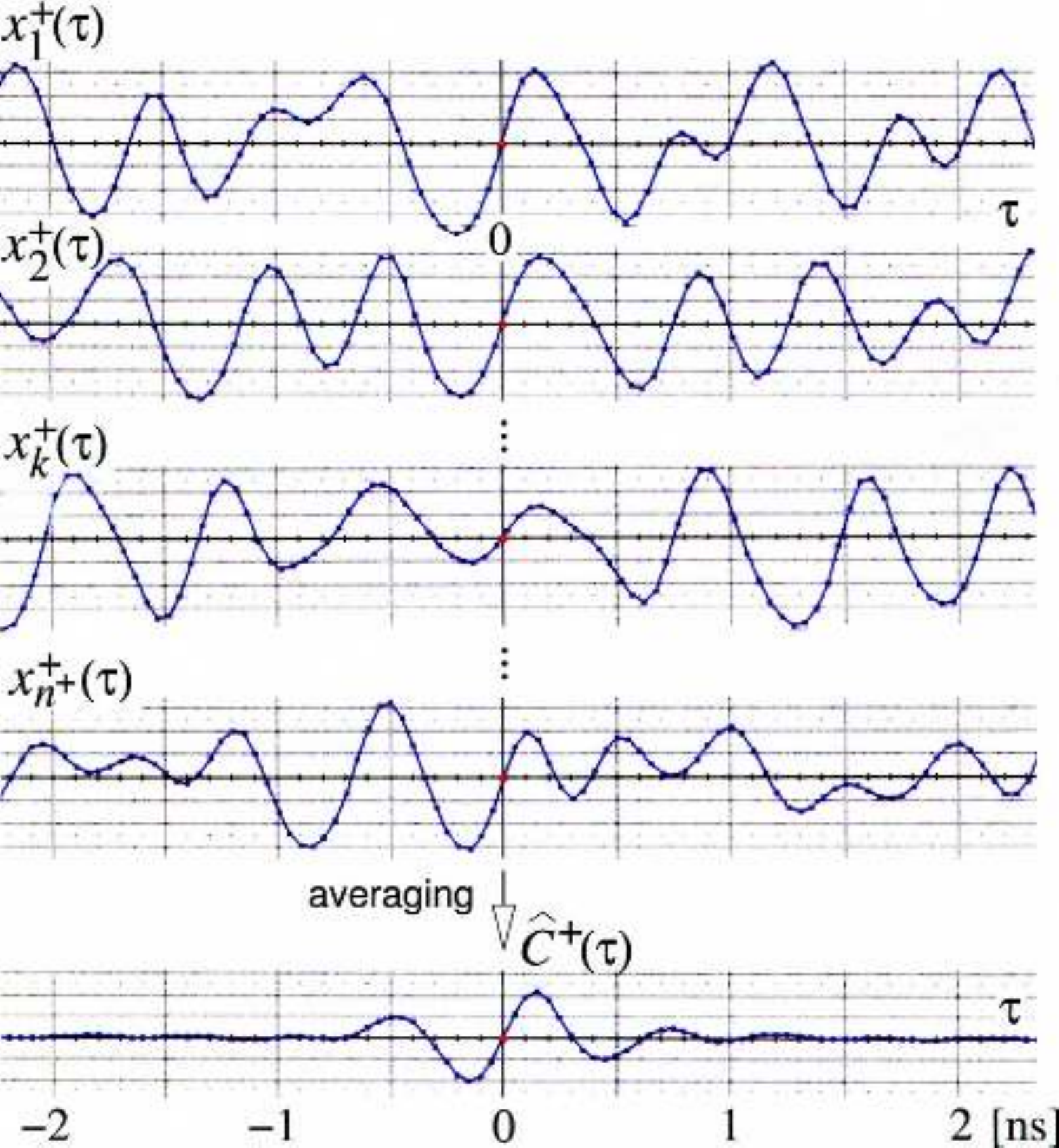}
	\vspace{-0.2cm}
	\caption{Crossjectories of wideband bandpass noise and their average.}
\end{figure}

For illustrative purposes, Fig.\,3 shows examples of crossjectories 
$\{x^+_k(\tau)\}$ obtained from a signal waveform $x(t)$ produced by a source of wideband noise with a 3-dB spectrum extending from $1$\,GHz to $2$\,GHz, as shown in Fig.\,4. In this experiment, discrete-time samples of the noise waveform $x(t)$ were taken at regular $50$\,ps-intervals. Fig.\,3 also shows an empirical zero-upcrossing interferogram, $\widehat{C}^+(\tau)$, obtained by averaging a large number $n^+$ of crossjectories $\{x^+_k(\tau)\}$. 

\begin{figure}[]
	\centering
	\includegraphics[width=7.cm]{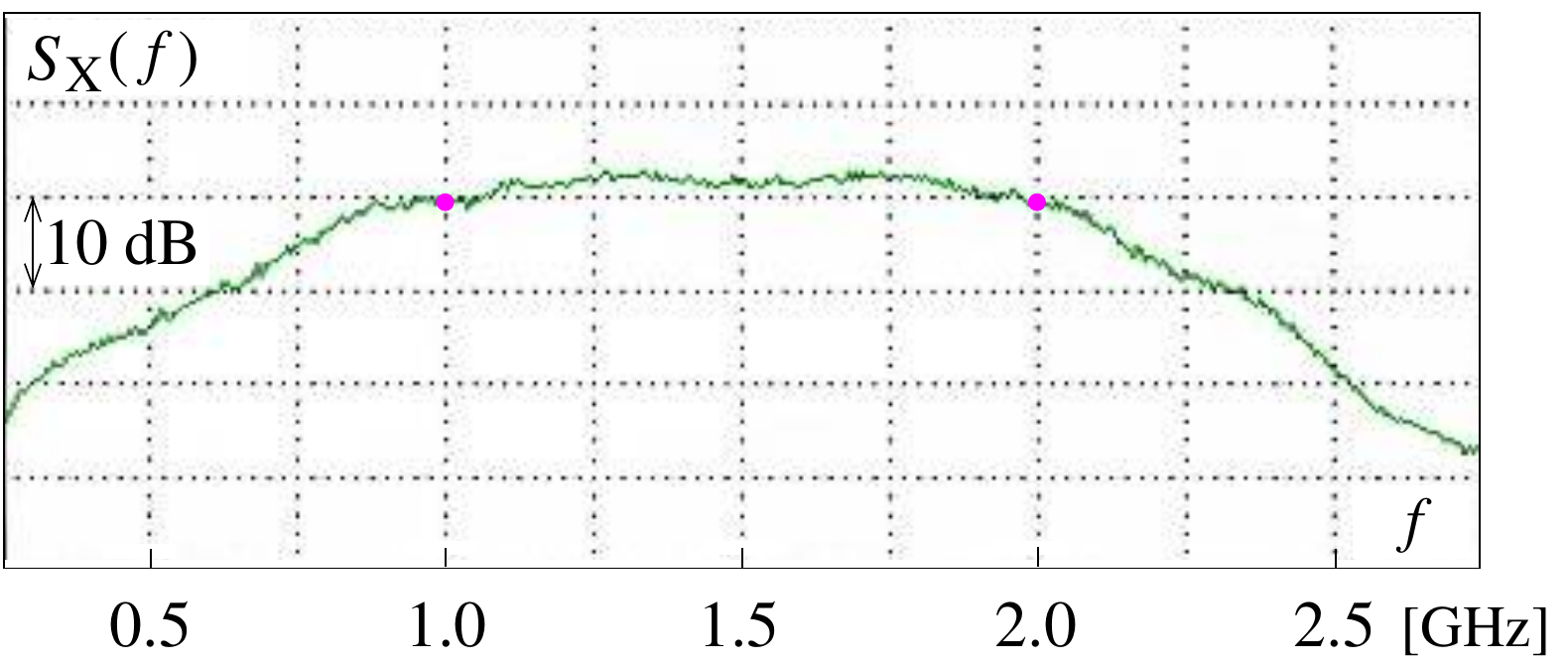}
	\vspace{-0.2cm}
	\caption{Power spectrum $S_X(f)$ of bandpass 
		noise used for experiments.}
\end{figure}

The modifications required to deal with zero {\emph {downcrossings}} are evident, since any zero downcrossing of $x(t)$ is a zero upcrossing of $-x(t)$. Assume that $n^-$ is the number of downcrossings occurring in a time interval $T$. Accordingly, the set $\{t_{\ell}\}$ of $n^-$ zero downcrossings will determine the set of $n^-$ corresponding crossjectories, $\{x^-_{\ell}(\tau) \triangleq x(t_{\ell}+\tau)\}$.
Consequently, the average
\begin{equation}
\widehat{C}^-(\tau) \, \triangleq \,
\frac {1}{n^-} \sum_{\ell=1}^{n^-}\mspace{-1mu} x_{\ell}^-(\tau) 
\end{equation}
being an estimate of the ensemble mean ${C}^-(\tau) \equiv \mathrm{E}\{X^-(\tau)\}$, is an empirical zero-downcrossing interferogram (6).

Crossjectories, $\{x(t_i+\tau); \,i=1,2,\ldots,n_c\}$,
associated with $n_c$ zero crossings (up or down) occurring at time instants $t_i$ can be used to determine the average
\begin{eqnarray}
\widehat{C}(\tau) &\!\!\!\triangleq\!\!\!& \frac {1}{n_c} \mspace{-1mu}
\sum_{i=1}^{n_c} (-1)^{\psi_i}x(t_i+\tau)    \\
&\!\!\!=\!\!\!& \frac {1}{n_c}\! \left[
\sum_{k=1}^{n^+}\mspace{-1mu} x^+_k(\tau) -
\sum_{\ell=1}^{n^-}\mspace{-1mu} x^-_{\ell}(\tau) \right] \nonumber
\end{eqnarray}
where $\psi_i=0$ for a zero upcrossing, and $\psi_i=1$ for a zero downcrossing. In the following, the average (11), resulting from a combination of two interferograms, (9) and (10), will be referred to as the {\em {empirical crosslation}} function.

\subsubsection*{Relationship Between Correlation and Crosslation}
By definition, a crossjectory associated with a zero crossing (up or down) at $t_i$ is the entire waveform $x(t)$ {\em {translated back}} by the 
time distance $t_i$. By using the translation property of the Dirac delta function, $\delta(t)$, respective crossjectories can be expressed as
\begin{eqnarray}
x_k^+(\tau)
&\!\!\!=\!\!\!&  \delta(t+t_k) \ast x(t)\nonumber \\
x_{\ell}^-(\tau) 
&\!\!\!=\!\!\!& \delta(t+t_{\ell}) \ast x(t)
\end{eqnarray}
where $\ast$ denotes convolution, defined by
\begin{equation}
(\eta \ast x)(\tau) \, \triangleq \, \int_{-\infty}^{+\infty}\!
\!\eta(\tau-\xi)\mspace{1mu}x(\xi)\,d\xi.
\end{equation}

Consequently, the empirical crosslation function (11) can be expressed in the following form
\begin{equation}
\widehat{C}(\tau) \, = \, \frac {1}{n_c}\, [d(t) \ast x(t)]
\end{equation}
where $d(t)$ is a train of bipolar delta pulses
\begin{equation}
d(t) \, \mspace{1mu}\triangleq\, \sum_{i=1}^{n_c}\mspace{1mu} (-1)^{\psi_i}  \delta(t+t_i). 
\end{equation}

The train $d(t)$ can be obtained from the underlying waveform $x(t)$ by performing a sequence of three operations:
\begin{IEEEeqnarray}{lll}
{\mbox{hard limiting:}}\quad &&v(t) \,=\, 
\mathrm{sgn}[x(t)] \nonumber \\
{\mbox{differentiation:}}\quad &&z(t) \,=\,\frac {1}{2}\, d\mspace{1mu}v(t)/d\mspace{1mu}t \nonumber \\
{\mbox{time reversal:}}\quad &&d(t) \,=\, z(-t) \quad
\end{IEEEeqnarray}
where 
\begin{equation}
\mathrm{sgn}[x(t)]\, \triangleq \,
\left\{
\begin{array}{r@{\,,\qquad}l}\,\,+1 & {x(t)>0}\\
                             \,\,0 & {x(t)=0}\\
                             \,\,-1 & {x(t)<0}\, .
\end{array} \right .
\end{equation}
Therefore, the empirical crosslation function (11) can be expressed as the convolution,
\begin{equation}
\widehat{C}(\tau) \, = \, \frac {1}{n_c}\mspace{1mu}
[z(-t) \ast x(t)].
\end{equation}

Since the number of zero crossings, $n_c$, is observed in a time interval $T$, the empirical crosslation function (11) assumes the following equivalent form,
\begin{equation}
\widehat{C}(\tau) \, = \, \frac {T}{n_c}\mspace{2mu} \widehat{R}_{ZX}(\tau) 
\end{equation}
where
\begin{equation}
\widehat{R}_{ZX}(\tau)\, \triangleq \, 
\left \langle z(t-\tau)\mspace{1mu}x(t) \right \rangle
\end{equation}
and the average is over the interval $T$, 

The average $\widehat{R}_{ZX}(\tau)$ will converge to the  cross-correlation function $R_{ZX}(\tau) \triangleq \mathrm{E}\{Z(t)X(t+\tau)\}$ as $T$ approaches infinity,
\begin{equation}
 \lim_{T \rightarrow \infty}\mspace{-1mu} \widehat{R}_{ZX}(\tau) \, = \,R_{ZX}(\tau).
\end{equation}
Hence, the empirical crosslation function $\widehat{C}(\tau)$ will converge to the mean ${C}(\tau)$,
\begin{equation}
{C}(\tau)\, = \, \frac {1} {\bar{n}_0}   \mspace{1mu}
 R_{ZX}(\tau) 
 \,= \, \frac{1}{2\mspace{1mu}\bar{n}_0}\mspace{1mu}  R'_{VX}(\tau)
\end{equation}
where $\bar{n}_0$ is the mean zero-crossing rate, i.e. the average number of zero crossings per unit time. In the following, the mean ${C}(\tau)$ will be referred to as the {\em {crosslation function}}.

Therefore, the crosslation function ${C}(\tau)$ of a random process $X(t)$ is proportional to the derivative of cross-correlation between a hard-limited version, $V(t) = \mathrm {sgn} [X(t)]$, of the process $X(t)$ and the process itself.

\subsection{Crosslation Function of a Separable Process}
When a zero-mean process $X(t)$ with variance $\sigma^2_X$ belongs to the class of {\emph {separable}} random processes [21], the cross-correlation function $R_{XV}(\tau)$ between the process $X(t)$ and its hard-limited version $V(t)$ can be expressed as
\begin{equation} 
R_{XV}(\tau) \, = \, \mu\mspace{1mu} R_{XX}(\tau)
\end{equation}
where $R_{XX}(\tau) \triangleq \mathrm{E}\{X(t)X(t+\tau)\} $ is the autocorrelation function of $X(t)$.
The constant $\mu$ of proportionality is given by [22]
\begin{equation} 
 \mu \, = \, \frac {1}{\sigma^2_X} 
 \int_{-\infty}^{\infty} \!\!|x|\mspace{2mu} p(x)\, dx
\end{equation}
where $p(x)$ is the probability density function of $X(t)$.

From the relationship, $R'_{VX}(\tau) = R'_{XV}(-\tau)$, and the fact that $R'_{XX}(\tau)$ is an odd function of $\tau$, it follows that $R_{VX}(\tau) = -\mu\mspace{1mu} R_{XX}(\tau)$. As a consequence, when $X(t)$ is a separable process, 
\begin{equation} 
C(\tau) \, = 
\, -\frac {\mu}{2\mspace{1mu}\bar{n}_0} \mspace{1mu} R'_{XX}(\tau)
\end{equation}
i.e. the crosslation function $C(\tau)$ of a separable process is proportional to the negative derivative of the autocorrelation function $R_{XX}(\tau)$ of the process.

According to the Wiener-Khintchin theorem, the autocorrelation function $R_{XX}(\tau)$ of a process $X(t)$ and its power spectral density $S_X(\omega)$, where $\omega$ is the angular frequency, form a Fourier pair. Therefore, the power spectral density $S_X(\omega)$ of a separable process can be determined from
\begin{equation} 
S_X(\omega) \, = \, \frac {2 j \mspace{1mu} \bar{n}_0}
{\mu \mspace{2mu} \omega} \mspace{2mu} {\mathcal {F}}\{C(\tau)\} 
\end{equation}
 where ${\mathcal {F}}\{\cdot\}$ denotes a Fourier transform. The formula (26) can be exploited in practice to develop a new approach to real-time spectral analysis of separable processes [24].

The class of separable random processes include [21]--[23]:\\
1. A Gaussian process, and also other elliptically symmetric processes;\\
2. A sine wave with phase or frequency modulation when the stationary modulation is independent of the carrier phase;\\
3. A signal comprising multiple sine waves, each having the same  amplitude and random phase distributed uniformly over a $(-\pi,\pi)$-interval;\\
4. A binary waveform alternating (randomly or otherwise) between two levels.\\
Furthermore, a product of two separable processes is also a separable process.
  
For example, in the case of a low-pass Gaussian process, 
\begin{equation} 
\mu \,=\, (1/\sigma_X)\sqrt{2/\pi } \quad \mbox{and} 
\quad \bar{n}_0 \,=\, B_X/\pi
\end{equation}
where $B_X$ is the root-mean-square (rms) bandwidth, defined by 
\begin{equation}
B_X \, \triangleq \,\, \left[
\frac {\int_{-\infty}^{\infty}\mspace{-2mu} \omega^2 S_X(\omega) \, d\mspace{1mu}\omega} {\int_{-\infty}^{\infty} S_X(\omega)\, 
	d\mspace{1mu}\omega} \right]^{\mspace{-2mu}1/2} \!\!.
\end{equation}
Also, $B_X = \sqrt{-R''_{XX}(0)/R_{XX}(0)}$, as a consequence of the Wiener-Khintchin theorem, .

Therefore, the crosslation function $C(\tau)$ of a Gaussian process can be expressed as
\begin{equation} 
C(\tau) \, = 
\, -\, \sqrt{\frac {\pi}{2}} \mspace{2mu}\frac{R'_{XX}(\tau)}{B_X \sigma_X} \,.
\end{equation}

\subsubsection*{The Slepian Process {\rm {[25], [26]}}}
When a waveform $x(t)$ is a realization of a zero-mean Gaussian process $X(t)$, the distribution of crossjectories $\{x^+(\tau)\}$ converges to the distribution of the so-called {\em {Slepian process}} $X_S^+(\tau)$, 
\begin{equation} 
X_S^+(\tau) \, = -\, U\mspace{1mu}
\frac{R'_{XX}(\tau)} {B_X^2\mspace{1mu} \sigma_X^2} 
\mspace{1mu}+\mspace{1mu} G(\tau); \quad \tau > 0
\end{equation}
where $U$ is a Rayleigh random variable (rv) with 
$\mathrm{E}\{U\}= \sqrt{\pi/2}\mspace{2mu}B_X \sigma_X$ and $\mathrm{E}\{U^2\}= 2\mspace{1mu}B_X^2\mspace{1mu} \sigma_X^2$.

The random process $G(\tau)$ is a zero-mean nonstationary Gaussian noise, independent of $U$, with the time-varying variance
\begin{equation}
	\sigma^2_{G}(\tau) \,=\, \sigma^2_X 
 - \frac{R^2_{XX}(\tau)} {\sigma_X^2} 
	\, - \, \frac{[R'_{XX}(\tau)]^2} {B_X^2\mspace{1mu} \sigma_X^2} 
\end{equation}
that increases from zero, at $\tau = 0$, to the limit 
$\sigma^2_X$, being reached at values of $\tau$ for which the effects of conditioning on a zero-crossing event have all vanished. 

The form (30) of the Slepian process shows that each crossjectory $x^+(\tau)$ is a randomly scaled replica of the negative derivative of the autocorrelation function $R_{XX}(\tau)$, perturbed by self-noise with intensity increasing from zero to $\sigma^2_X$. Since a downcrossing of $X(t)$ is an upcrossing of $-X(t)$, this observation also applies to crossjectories associated with downcrossings. Also, since a Gaussian process is time reversible, and an upcrossing of $X(t)$ is also an upcrossing of $-\!X(-t)$, the model (30) is valid for $\tau < 0$.

\begin{figure}[]
	\centering
	\includegraphics[width=5.9cm]{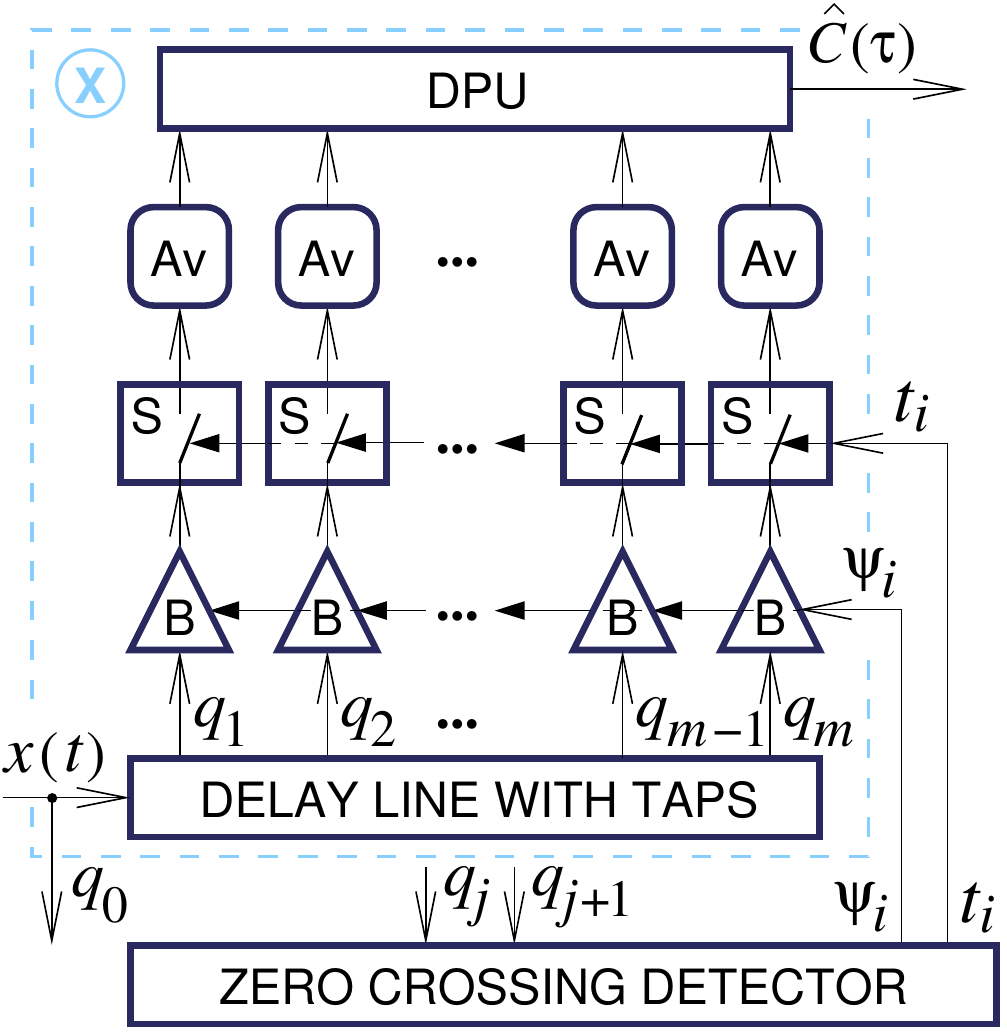}
	\vspace{-0.2cm}
	\caption{Functional block diagram of a crosslator.}
\end{figure}

\subsection{Crosslator}
Fig. 5 is a functional block diagram of a crosslator [27] -- a real-time system designed to determine an empirical crosslation function $\widehat{C}(\tau)$ of a waveform. The crosslator consists of two main functional blocks: a crosslator subsystem \textcircled{\scriptsize {\sf{X}}} and a zero-crossing detector.

A waveform $x(t)$ under analysis is applied to a delay line with 
$m$ taps. Each of the taps provides a time-delayed replica of the waveform $x(t)$, and at any time instant, the values observed at the $m$ taps of the delay line form jointly a representation of the waveform $x(t)$ propagating along the line.

The ascending order of the tap subscripts, 
$1,2,\ldots,m\!-\!1,m$ corresponds to an increasing amount of {\em {elapsed time}}. The delay line can only contain past data, and the input, a notional tap $q_0$, represents an ever-changing present moment in time. 

An event of crossing a zero level by the waveform $x(t)$ is detected, when the outputs of two closely-spaced taps, $q_j$ and $q_{j+1}$, have opposite signs. A zero upcrossing occurring in {\emph {relative}} time will be declared, when the output of $q_j$ is positive; otherwise, the detected zero crossing will be a downcrossing.
When the zero-crossing detector uses taps $q_j$ and $q_{j+1}$, a nominal zero crossing will appear to be 'located' in the middle, between these two taps.

\begin{figure}[]
	\centering
	\includegraphics[width=8cm]{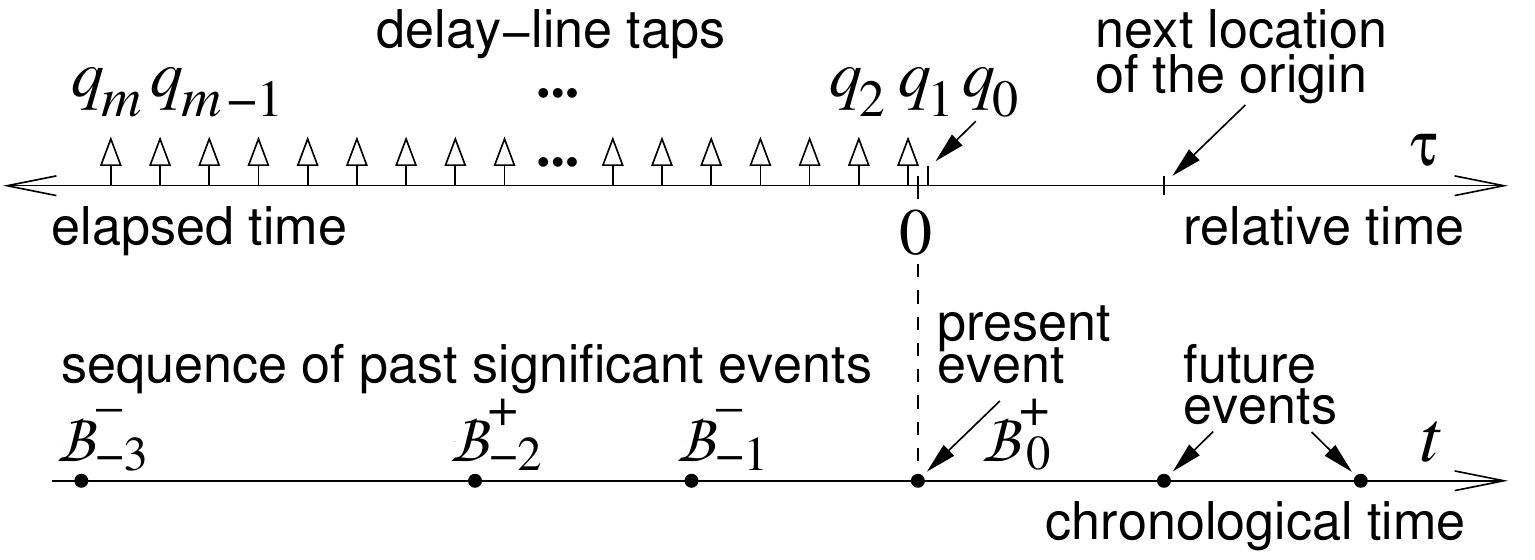}
	\vspace{-0.2cm}
	\caption{Chronological time, elapsed time and relative time.}
\end{figure}

Fig. 6 shows relationships between chronological time $t$, relative time $\tau$ and elapsed time. A sequence of significant events, 
$\mathcal{B}_{-3}^-,\mathcal{B}_{-2}^+,\mathcal{B}_{-1}^-,\mathcal{B}_0^+$, 
of zero crossings, is marked by dots on the axis of chronological time $t$. 
Each new zero crossing, detected in chronological time, will define a new location of the origin of elapsed time. Consequently, the origin of elapsed time (and also that of relative time $\tau$) is moving in steps along the axis of chronological time, each new location being determined by a next event of zero crossing. 

For example, in Fig.\,6, the origin of elapsed time is determined by an upcrossing $\mathcal{B}_0^+$, detected between taps $q_0$ and $q_1$. In the  (chronological) past, the origin of elapsed time had coincided with the events: $\mathcal{B}_{-1}^-$ (immediate past), $\mathcal{B}_{-2}^+$ and $\mathcal{B}_{-3}^-$. In the uncertain (chronological) future, the origin of elapsed time will be shifted to the locations of hypothetical future events of zero crossings.

\begin{figure}[]
	\centering
	\includegraphics[width=7.5cm]{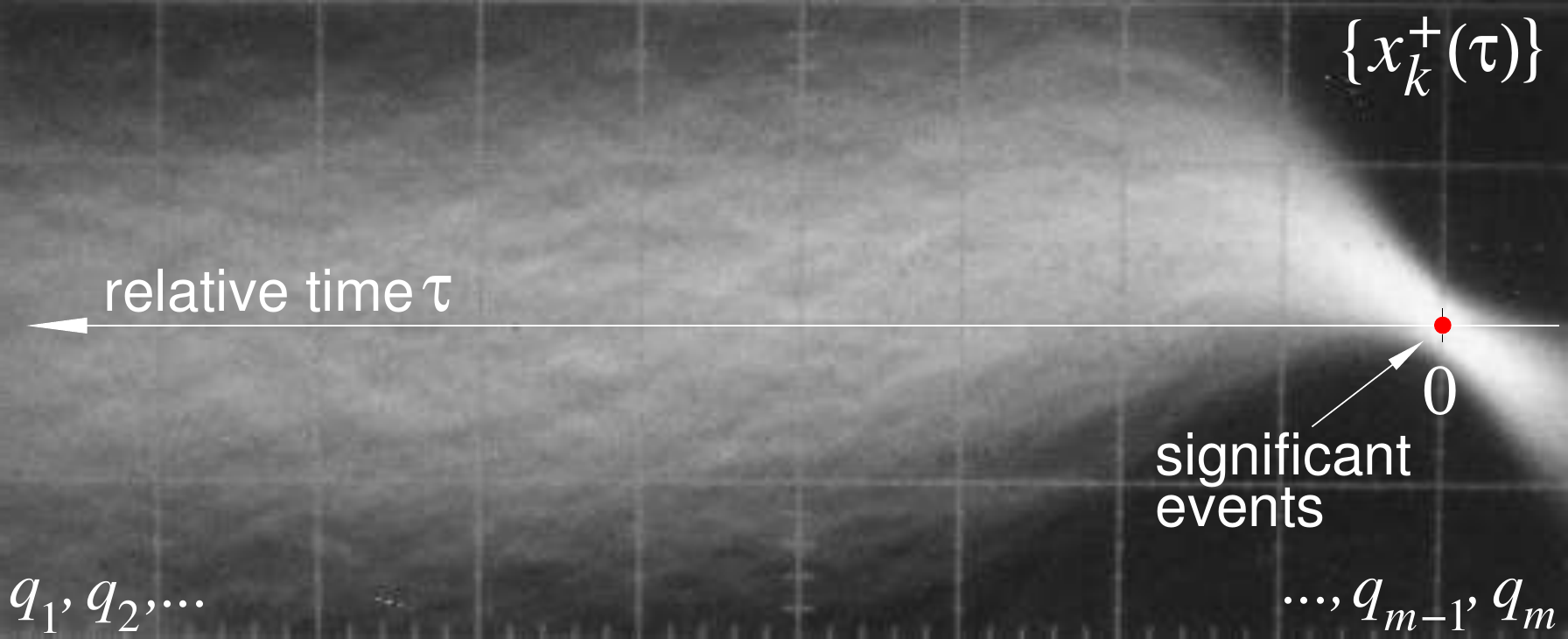}
	\vspace{-0.2cm}
	\caption{Observed crossjectories when zero crossings are detected between taps $q_{m-1}$ and $q_m$.}
\end{figure}

In the crosslator of Fig.\,5, the delay line can contain continually-changing past data, and only a half of each crossjectory $x_i(\tau)$, for $\tau <0$, can be processed in real time. However, the other half of each crossjectory, for $\tau > 0$, can also be processed, when the zero-crossing detector uses taps $q_{m-1}$ and $q_m$ to shift the origin of elapsed time. Crossjectories obtained in such a 'future-in-the-past' arrangement are shown in Fig.\,7; they correspond, in reverse time, to crossjectories of Fig.\,2. An optimum choice of the length of the delay line as well as the positions of adjacent taps to be used by the zero-crossing detector will depend on the intended application of the crosslator.

The zero-crossing detector supplies, at its two outputs, timings $t_i$ and types $\psi_i$ of detected zero crossings. When $\psi_i=0$, each of the taps $q_j$,  $j=1,2, \ldots, m$ is directly connected via a buffer/inverter {\sf {B}} to a respective sampler {\sf {S}}; however, when $\psi_i= 1$, the buffer/inverter {\sf {B}} changes the polarity of the tap prior to connecting it to the sampler {\sf {S}}.

At the time instants $\{t_i\}$, discrete-time samples are transferred to the bank of averaging circuits {\sf {Av}}, arranged to operate over a fixed time interval or in a recursive mode. The resulting values of $m$ averages are then supplied to the data processing unit {\sf {DPU}} for further processing, such as index reversal and interpolation, to produce an empirical crosslation function $\widehat{C}(\tau)$ of relative time $\tau$.

\section{Complex Crosslation Function}
The shape of the crosslation function $C(\tau)$ is well suited to some applications, such as time-delay tracking or bandwidth estimation [27]--[30]. Also, since the function $C(\tau)$ is related to the conditional mean of a local increment, it can be used to construct an  absolute local structure function, $|C(\tau)|$, with a cusp at $\tau=0$. However, in other applications, such as multipath discrimination or signal detection in noise, it would be useful to have a complementary representation of the crosslation function $C(\tau)$, in which a prominent peak will appear at the origin $\tau=0$ of relative time. 

In the following, such a representation of the crosslation function $C(\tau)$ will be determined for the class of separable processes. The complementary representation will then be used in conjunction with the underlying crosslation function to construct a {\emph {complex}} crosslation function.

\subsection{Autoference ${\mspace{-5mu}}^3$ Function of a Separable Process}
A crosslation function $C(\tau)$ of a separable process $X(t)$ is an odd function of $\tau$. A corresponding even function, $A(\tau)$, with the same power spectrum density, yet having a positive peak at $\tau=0$, can be obtained by applying an inverse Hilbert transform [31, eq. (1.10)] to $C(\tau)$,
\begin{equation}
A(\tau)  \,\triangleq \,\mathcal{H}^{-1}\{C(\tau)\} \,= \,
-C(\tau) \ast  \left[ 1/(\pi t) \right]
\end{equation}
where $\mathcal{H}^{-1}\{\cdot\}$ denotes an inverse Hilbert transform. In the following, the even function $A(\tau)$ will be referred to as the {\emph {autoference}}{\footnote {the name selected for the proposed new function.}} function.

Since [31, eq. (1.52)]
\begin{equation}
\mathcal{F} \! \left \{ 1/(\pi\, \tau)        \right\}  \,=\,-j\mspace{2mu} \mathrm{sgn}\{\omega \}, 
\end{equation}
where $j^2 = -1$,
the autoference function $A(\tau)$ can also be determined from
\begin{equation}
A(\tau)  \,=\,  \mathcal{F}^{-1}\mspace{-1mu} \left \{
   j\mspace{2mu} \mathrm{sgn}\{\omega \} \mspace{1mu} \mathcal{F}\{C(\tau)\}
\right\} .
\end{equation}

From (25) and the Wiener-Khinchin theorem it follows that a crosslation function $C(\tau)$ of a separable random process $X(t)$ can be expressed as
\begin{equation}
C(\tau)  \,=\,
- \frac {\mu}{2\mspace{1mu}\bar{n}_0}\, \mathcal{F}^{-1} 
\{j\mspace{1mu}\omega\mspace{1mu} S_X(\omega)\} .
\end{equation}
Finally, the autoference function $A(\tau)$ can be determined from
\begin{equation}
A(\tau)  \,=\, \frac {\mu}{2\mspace{1mu}\bar{n}_0} \mspace{2mu} \mathcal{F}^{-1} \mspace{-2mu}
\left \{|\omega|\mspace{1mu} S_X(\omega)  \right \}
\end{equation}
where $\mu$ is given by (24) and $\bar{n}_0$ is the mean zero-crossing rate.

Therefore, in crosslation-based signal processing, 
higher frequencies are accentuated, whereas lower ones are attenuated. 
As a result, the rms bandwidth of a waveform will generally be increased; 
 also, in the case of a bandpass waveform, 
 its centre frequency will be shifted to a higher value.
This phenomenon of {\em {virtual high-frequency emphasis}} 
can be exploited to design optimal waveforms for crosslation-based signal processing.

An empirical autoference function $\widehat{A}(\tau)$ can be obtained directly from the crosslation $\widehat{C}(\tau)$ by exploiting (32). However, $\widehat{A}(\tau)$ can also be determined by averaging Hilbert-transformed crossjectories in accordance with (11), modified as follows
\begin{eqnarray}
\widehat{A}(\tau) &\!\!\!=\!\!\!& \frac {1}{n_c}\mspace{1mu}
\mathcal{H}^{-1}\! \left \{
\sum_{i=1}^{n_c} (-1)^{\psi_i}x(t_i + \tau) \right \} \nonumber \\
 &\!\!\!=\!\!\!& \frac {1}{n_c}\mspace{-1mu}
\sum_{i=1}^{n_c} (-1)^{\psi_i}y(t_i + \tau) 
\end{eqnarray}
where $y(t_i + \tau) = \mathcal{H}^{-1}\{x(t_i + \tau)\}$. The above approach is especially well-suited to processing of bandpass waveforms, when a required Hilbert transformer can be replaced by a quadrature phase splitter. 

\begin{figure}[]
	\centering
	\includegraphics[width=6.9cm]{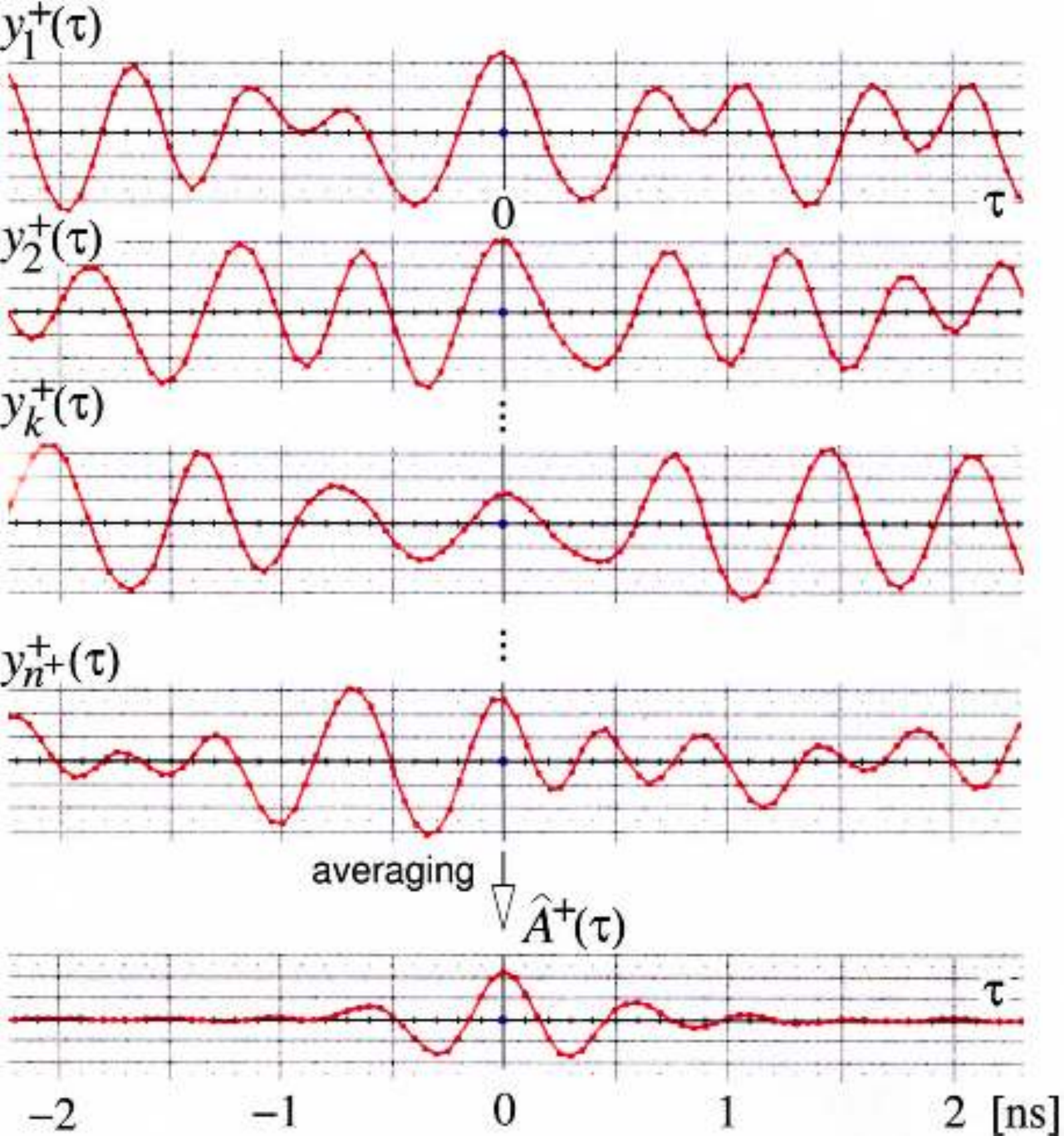}
	\vspace{-0.2cm}
	\caption{Hilbert-transformed crossjectories of Fig.\,3 and their average.}
\end{figure}
As an illustration of the method, Fig.\,8 shows trajectories $\{y^+_k(\tau)\}$ obtained from the corresponding crossjectories $\{x^+_k(\tau)\}$ of Fig.\,3  by passing them through a wideband quadrature phase splitter. As more and more trajectories $\{y^+_k(\tau)\}$ are being combined together, a distinct interference pattern, akin to a wave packet [32], will begin to emerge.

Fig.\,8 also shows an empirical autoference function $\widehat{A}^+(\tau)$ that has resulted from averaging a large number $n^+$ of trajectories $\{y^+_k(\tau)\}$. The function $\widehat{A}^+(\tau)$ can be regarded as a specific zero-crossing waveform interferogram in which zero crossings are determined from a 'Hilbert image' of a waveform whose trajectories are being averaged.

The autoference function $A(\tau)$ is represented by a first-order statistic (37), and as such is expressed in units of the waveform's amplitude (e.g., Volts). In particular, the quantity $A(0)$ is the  mean of $Y(t)=\mathcal{H}^{-1}\{X(t)\}$, conditioned on zero crossings of the underlying process $X(t)$. 

The autoference function can be used to localize in time a wideband waveform $x(t)$ of extended (theoretically infinite) duration with a resolution depending on the shape of a modified spectrum $|\omega|S_X(\omega)$. Therefore, in applications related to object localisation and imaging, an empirical autoference function $\widehat{A}(\tau)$ can be used instead of empirical correlation function, $\widehat{R}_{XX}(\tau)$, given by (3).

\subsubsection*{Autoference as Correlation}
A useful interpretation of the autoference function is obtained when (37) is expressed as
\begin{equation}
\widehat{A}(\tau) \,=\,
 \frac {1}{n_c}\mspace{-1mu}
\sum_{i=1}^{n_c} \mspace{-1mu}
\big ({\mathrm{sgn}}[y(t_i)] \big )\mspace{2mu} y(t_i + \tau) 
\end{equation}
where the time instants $\{t_i\}$ are determined from zero crossings of $x(t) = \mathcal{H}\{y(t)\}$. The average (38) is then an empirical {\emph {cross-correlation}} function between two waveforms: a hard-limited version of the waveform $y(t)$ and the waveform itself; however, the  time instants $\{t_i\}$ are obtained from zero crossings of the Hilbert transform of $y(t)$. 
The form (38) may lead to a simpler implementation, since a Hilbert-transformed waveform, $x(t) = \mathcal{H}\{y(t)\}$, is only used to determine timings of its zero crossings.

In some applications, such as detection of signals in noise, it may be  advantageous to use an empirical {\emph {weighted}} autoference function, defined by
\begin{equation}
\widehat{A}_w(\tau) \, \triangleq \, \frac {1}{n_c} \mspace{-1mu}
\sum_{i=1}^{n_c} y(t_i)\mspace{2mu} y(t_i + \tau).
\end{equation} 
The average (39) is simply an empirical autocorrelation function of a waveform $y(t)$, where the time instants $\{t_i\}$ are determined from zero crossings of the Hilbert transform, $\mathcal{H}\{y(t)\}$, of the waveform being processed. The weighted autoference function (38) can be used instead of empirical correlation function, $\widehat{R}_{XX}(\tau)$, given by (3).

In active echolocation/imaging applications, the averages (38) and (39) can be made equivalent by designing the waveform $y(t)$ so that the values $\{y(t_i)\}$ are all equal. One solution is to employ, as an illuminating waveform, a sinusoidal wave with suitable phase/frequency modulation. 

\subsubsection*{Bandpass Filtering}
For efficient crosslation-based processing of a waveform, observed over a time interval $T$, the number $n_c$ of zero crossings (hence, the number of samples being averaged), should be equal to the {\em {number of degrees of freedom}}, $\Lambda$, i.e. the number of {\emph {uncorrelated}} samples, characterising the waveform. When $n_c < \Lambda$, the waveform is said to be {\emph {underdetermined}} by its zero crossings, whereas when $n_c > \Lambda$, the waveform will become {\emph {overdetermined}}.
 
For a low-pass Gaussian process $X(t)$, the expected number $n_c$ of zero crossings, occurring in a time interval $T$, is given by Rice's formula, $n_c = TB_X/\pi$, where $B_X$ is the rms bandwidth (28) of the process. For large $T$, the approximate number of degrees of freedom can be determined from [33]
\begin{equation}
\Lambda \,\,= \,\,\frac {T\mspace{-1mu} R^2_{XX}(0)}{ \int_{-\infty}^{\infty}\! 
R^2_{XX}(\tau)\, d\tau}
\end{equation}
where $R_{XX}(\tau)$ is the autocorrelation function of $X(t)$. 

In a special case of a Gaussian process with a uniform spectrum that vanishes for $|\omega| > W$, the number $\Lambda$ of degrees of freedom is simply $WT/\pi$, whereas the rms bandwidth equals $W/\sqrt{3}$. Therefore, the number,
$WT/(\pi \sqrt{3})$ of zero crossings provides only a fraction, $1/\sqrt{3} \approx 0.58$, of the number of degrees of freedom, $\Lambda = WT/\pi$, required for efficient processing.

 In a band-limited bandpass process, with the spectrum uniform in the frequency band, $W_1 < |\omega| < W_2$, the number of zero crossings will be equal to the number,  $\Lambda = (W_2 - W_1)T/\pi$, of degrees of freedom, when [34]
\begin{equation}
W_2/W_1 \, \,\le \,\,    (7+\sqrt{33}\mspace{2mu} )/4\,\,
\approx \, \, 3.19.
\end{equation}
It can be shown [35] that when the spectrum of a Gaussian process $X(t)$ has a Gaussian shape, $\exp[-\omega^2/(2B^2_X)]$, the number of degrees of freedom is 
\begin{equation}
\Lambda  \,= \,   B_x\mspace{1mu} T /\sqrt{\pi}.
\end{equation}
Since $n_c=B_x T/ \pi$, zero crossings can only provide a fraction, $1/\sqrt{\pi} \approx 0.56$, of the required number (42) of degrees of freedom.
In order to satisfy the inequality, $n_c \ge \Lambda$, it has been proposed to employ a bank of bandpass filters, each with an impulse response being a difference between two Gaussian functions with a constant ratio (of about $3$) of their respective bandwidths [35]. 

The above analysis demonstrates that in the case of a low-pass Gaussian random waveform, having a Gaussian or band-limited uniform spectrum, efficient processing can be achieved, when a bank of parallel bandpass filters is used to decompose the spectrum of the waveform into a number of frequency bands, each approximately $1.5$-octave wide. A zero-crossing interferogram of the waveform is then obtained by combining individual interferograms, each determined for a separate frequency band. 

However, when a waveform has an extremely slowly decaying spectrum, such as  Lorentzian (Section IV.\,C), zero crossings may even overdetermine the waveform.
Moreover, a product of random or pseudorandom waveforms with predetermined characteristics can be used to obtain a specified number of zero crossings [36]. 

\subsection{Representations of a Complex Crosslation Function}
A complex crosslation function $\mathcal{A}(\tau)$ is defined by
\begin{equation}
\mathcal{A}(\tau) \,\, \triangleq \,\, A(\tau) + j\mspace{1mu}C(\tau)
 \,=\, A(\tau) + j\mspace{1mu}\mathcal{H}\{A(\tau)\}.
\end{equation}
The function $\mathcal{A}(\tau)$ can be represented by a spatial curve, a {\em {crosslation helix}}, having two orthogonal components, $A(\tau)$ and $C(\tau)$. Since the real and imaginary parts, $A(\tau)$ and $C(\tau)$, of a complex crosslation function $\mathcal{A}(\tau)$ are equal to respective conditional means, each of them has units of the amplitude of a waveform being processed.

\begin{figure}[]
	\centering
	\includegraphics[width=8.5cm]{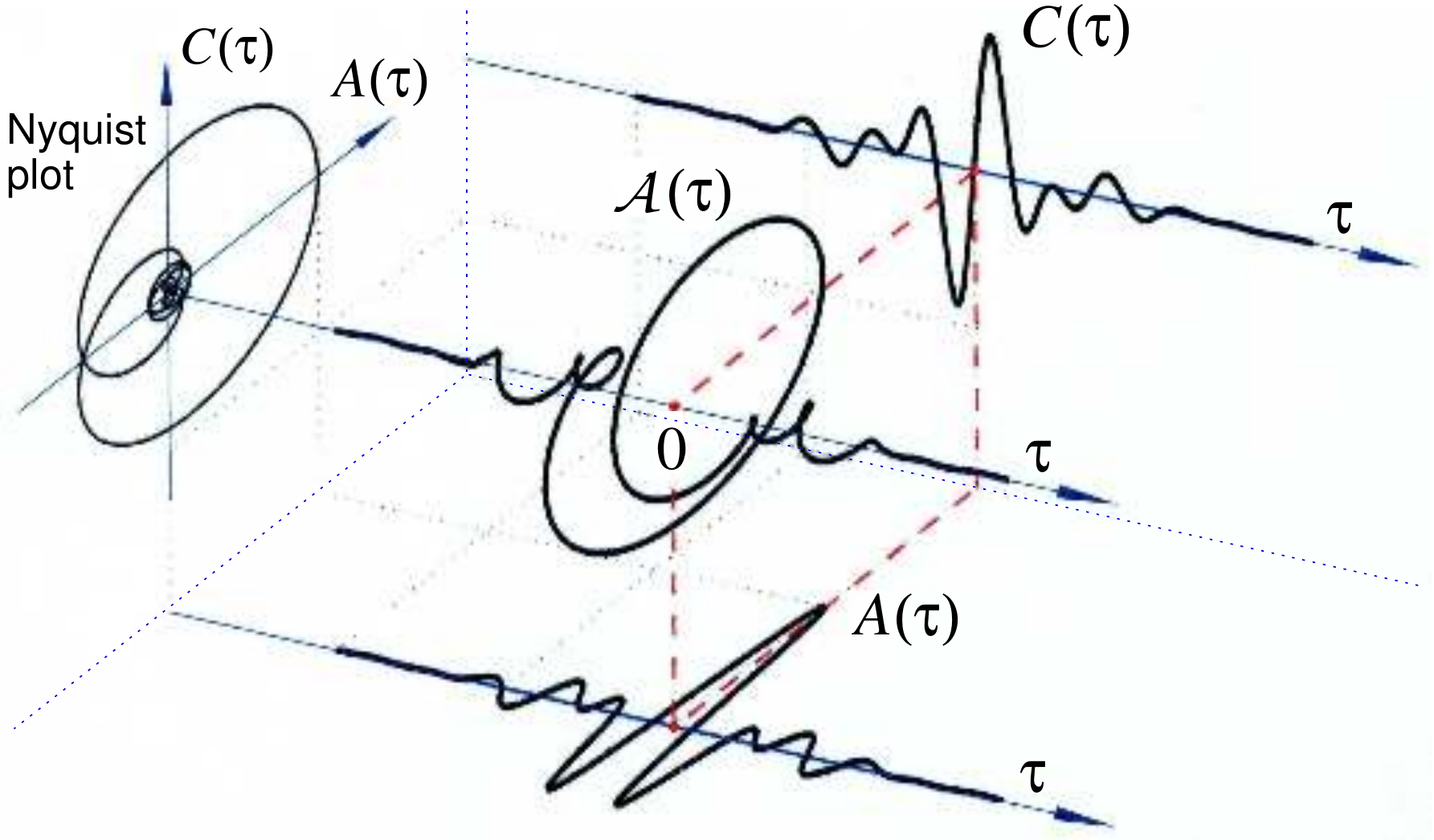}
	\vspace{-0.1cm}
	\caption{Examples of a crosslation helix $\mathcal{A}(\tau)$, its two components: $A(\tau)$ and $C(\tau)$, and a Nyquist plot.}
\end{figure}

Components of a complex crosslation function $\mathcal{A}(\tau)$ can be used to define the crosslation envelope as follows
\begin{equation}
|\mathcal{A}(\tau)| \, \triangleq \,
\sqrt{A^2(\tau) + C^2(\tau) }.
\end{equation}
The envelope (44) is employed for waveform localisation, when in addition to an unknown time shift of a crosslation helix, its angle of rotation ('phase') is also unknown.

In some applications, it may be of interest to exploit the Nyquist plot, i.e. a parametric plot of the imaginary part $C(\tau)$ versus the real part $A(\tau)$, as the relative time $\tau$ is varied from minus to plus infinity. The Nyquist plot can be viewed as a specific form of a waveform 'fingerprint'; accordingly, the plot may be used for waveform classification and discrimination.

Fig.\,9 is an example of a crosslation helix $\mathcal{A}(\tau)$ and its components, $A(\tau)$ and $C(\tau)$, determined for a multi-sine waveform. A corresponding Nyquist plot, $C(\tau)$ versus $A(\tau)$, is also shown as a closed planar curve.

\begin{figure}[]
	\centering
	\includegraphics[width=7.5cm]{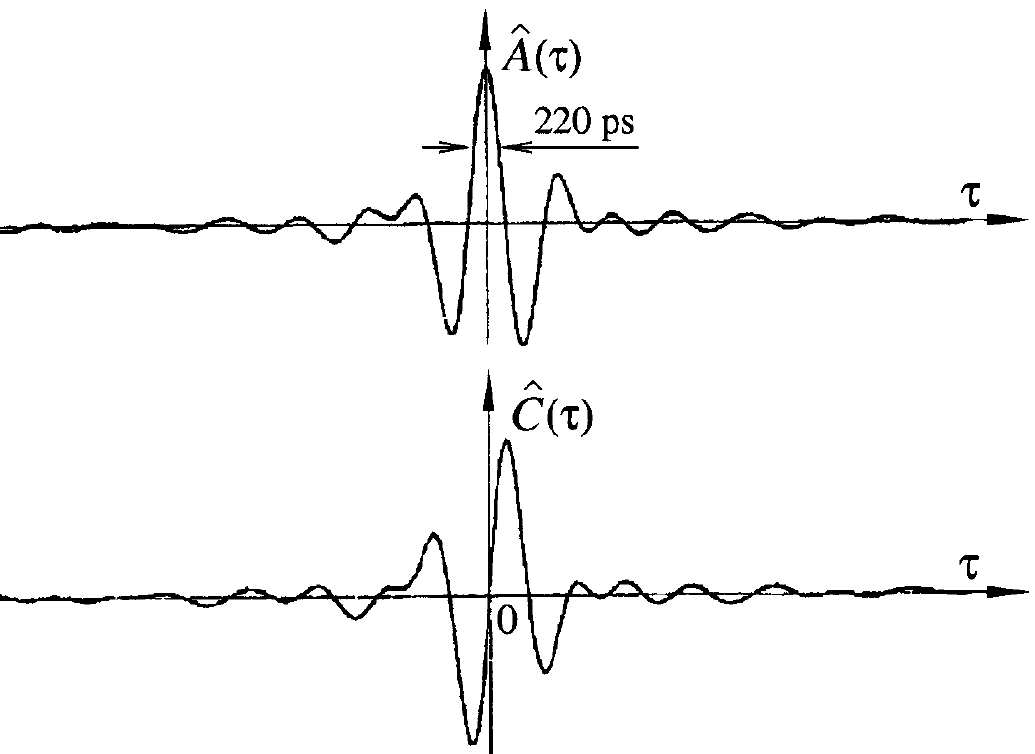}
	\vspace{-0.0cm}
	\caption{Components of a complex crosslation function of wideband noise with a spectrum extending from $1$\,GHz to $2$\,GHz.}
\end{figure}

Fig.\,10 shows two orthogonal components, $\widehat{A}(\tau)$ and $\widehat{C}(\tau)$, of a complex crosslation function, $\widehat{\mathcal{A}}(\tau)$, obtained experimentally for a Gaussian noise having a power spectrum extending from $1$\,GHz to $2$\,GHz. The experimental data have confirmed that the nominal centre frequency of $1.5$\,GHz has been shifted to $1.58$\,GHz, as a result of the phenomenon of virtual high-frequency emphasis. 

\begin{figure}[]
	\centering
	\includegraphics[width=6.8cm]{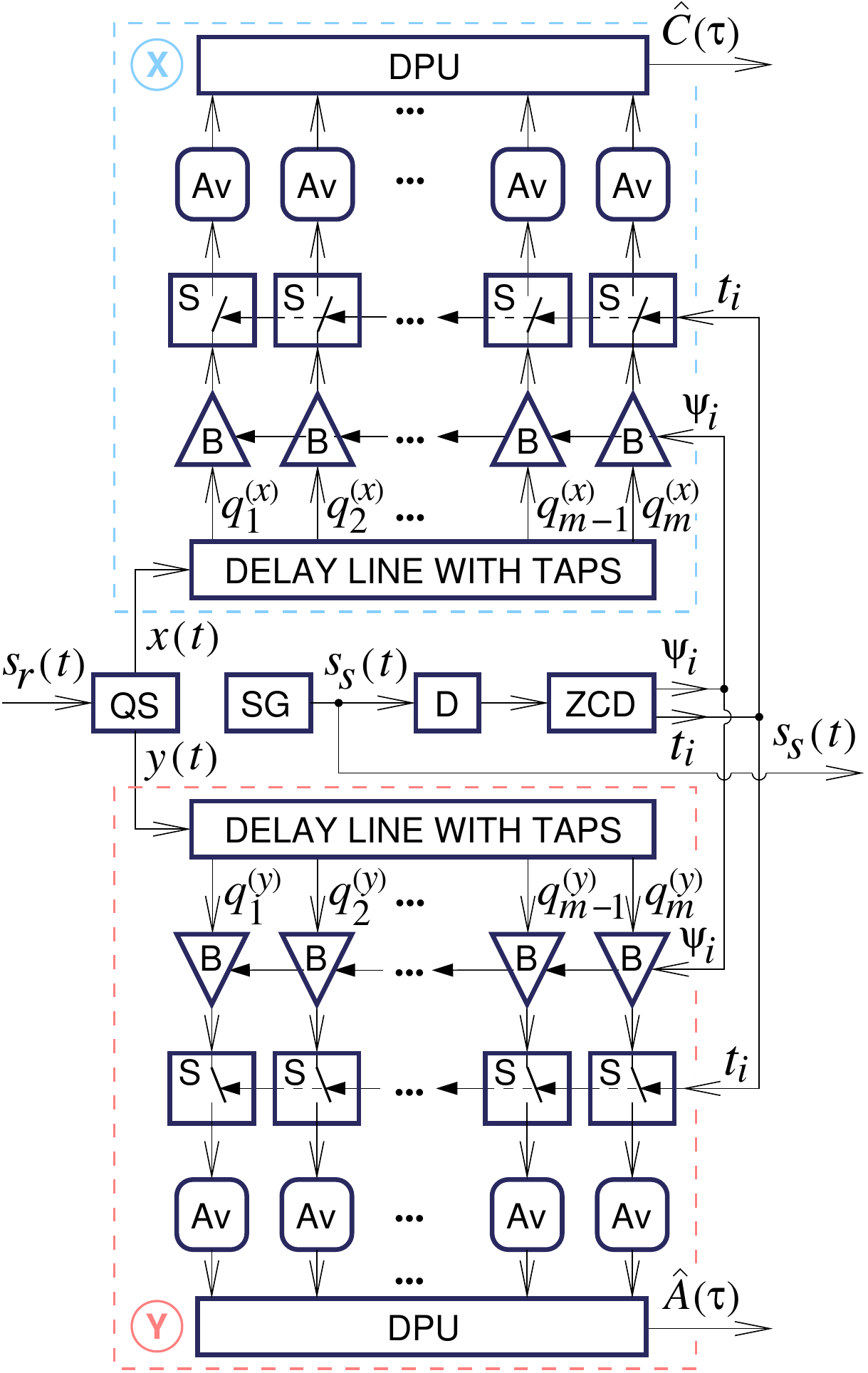}
	\vspace{-0.2cm}
	\caption{Block diagram of a conceptual echolocation system incorporating a complex crosslator.}
\end{figure}

\subsection{Crosslation-Based Echolocation System}
A block diagram of a conceptual echolocation system employing a complex crosslator is shown in Fig.\,11. A suitable interrogating signal $s_s(t)$, produced by a signal generator {\small {\sf {SG}}}, is used to illuminate some region of interest. The signal $s_s(t)$, delayed by a constant delay line {\small {\sf {D}}}, is applied to a zero-crossing detector {\small {\sf {ZCD}}} whose two outputs supply timings $t_i$ and types $\psi_i$ of detected zero crossings.
The parameters, $t_i$ and $\psi_i$, are used to control the operations of two identical crosslator subsystems, 
\textcircled{\scriptsize {\sf{X}}} and 
\textcircled{\scriptsize {\sf{Y}}}, each having the same configuration as the subsystem \textcircled{\scriptsize {\sf{X}}} of Fig.\,5. 

A signal $s_r(t)$, reflected by objects present in the illuminated region, is applied to the input of a quadrature splitter {\small {\sf {QS}}} that produces at its outputs two representations, $x(t)$ and $y(t)$, of $s_r(t)$, where $x(t) = \mathcal{H}\{y(t)\}$. The two waveforms, $x(t)$ and $y(t)$, are processed in respective crosslator subsystems, 
\textcircled{\scriptsize {\sf{X}}} and 
\textcircled{\scriptsize {\sf{Y}}}, to obtain two empirical functions,  
$\widehat{C}(\tau)$ and $\widehat{A}(\tau)$. The functions are then used to determine the crosslation envelope $|\widehat{\mathcal{A}}(\tau)|$, defined by (44). The envelope $|\widehat{\mathcal{A}}(\tau)|$ will provide information about locations and reflectivity of various objects illuminated by the transmitted signal $s_s(t)$.

Fig.\,12 shows the results of an experiment in which an illuminating signal is 
a $1.4$\,GHz sinusoidal carrier modulated in frequency by Gaussian noise; in this case, the resulting bandwidth is equal to approximately $450$\,MHz.
The two components, $\widehat{A}(\tau)$ and $\widehat{C}(\tau)$, of a complex crosslation function of a reflected signal have been used to determine the envelope $|\widehat{\mathcal{A}}(\tau)|$, and also the time resolution 
($\sim 1.9$\,ns) . 

\begin{figure}[]
	\centering
	\includegraphics[width=7.5cm]{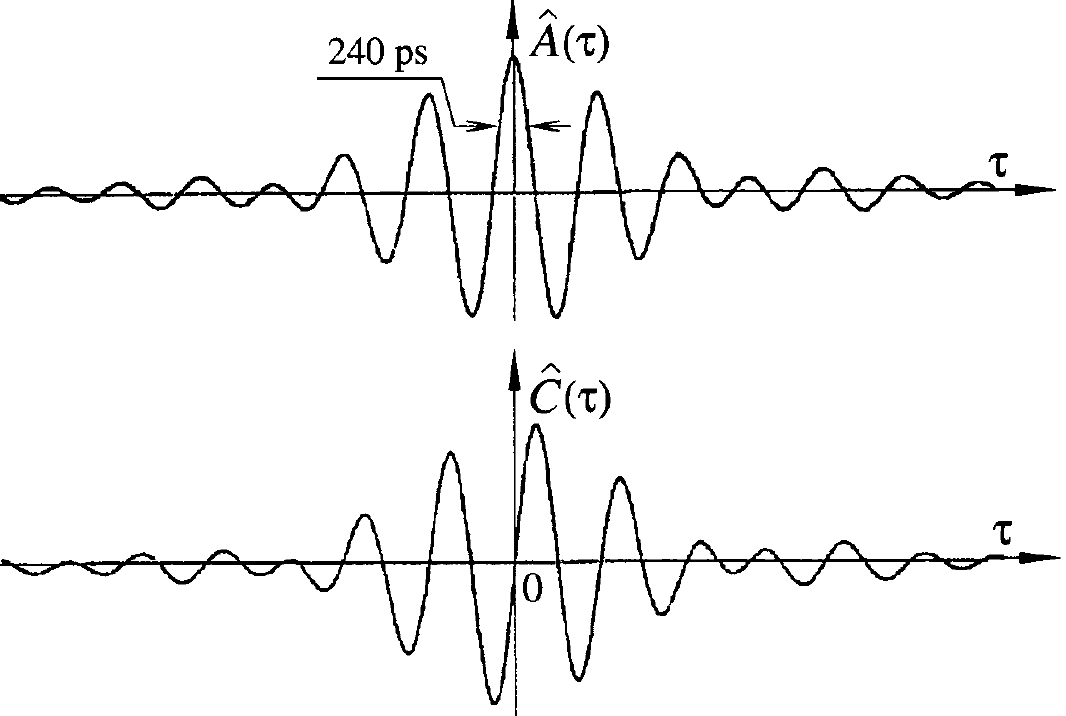}
	\vspace{-0.0cm}
	\caption{Empirical autoference, $\widehat{A}(\tau)$, and crosslation,  $\widehat{C}(\tau)$, functions of a delayed replica of a $1.4$\,GHz sinusoidal carrier modulated in frequency by Gaussian noise.}
\end{figure}

\section{Localisation of Waveforms in Time}
The performance of echolocation and imaging systems depends on their ability to resolve time-shifted replicas of a waveform. In conventional correlation-based systems, the potential time resolution associated with a random process $X(t)$ can be determined from the envelope of its complex autocorrelation function $\mathcal{R}(\tau)$. 

The complex autocorrelation function $\mathcal{R}(\tau)$ of 
a random process $X(t)$ is defined by
\begin{equation}
\mathcal{R}(\tau) \,\,\triangleq \,\, {\frac{1}{2} }\left [
\mathrm{E}\{\mathcal{X}^*(t)\mspace{1mu}\mathcal{X}(t+\tau)\} \right ]
\end{equation}
where
\begin{equation}
\mathcal{X}(t) \,\,\triangleq\,\, X(t) + j\mspace{1mu}Y(t)
\end{equation}
is an analytic random process whose imaginary part, $Y(t)$, is a Hilbert transform of $X(t)$; $Y(t) \triangleq \mathcal{H}\{X(t)\}$. The  power spectrum density of the analytic process $\mathcal{X}(t)$ is 
$2S_X(\omega)$, when $\omega > 0$, and zero for all negative frequencies, $\omega < 0$.

The complex autocorrelation function $\mathcal{R}(\tau)$ 
can be expressed as [31, eq. (7.112)]
\begin{equation}
\mathcal{R}(\tau) \,\,=\,\,  R_{XX}(\tau) + j\mspace{1mu}R_{XY}(\tau)
\end{equation}
where
\begin{equation}
R_{XY}(\tau) \,\,\triangleq \,\,
\mathrm{E}\{X(t)\mspace{1mu}Y(t+\tau)\} \, =\, \mathcal{H}\{R_{XX}(\tau)\}.
\end{equation}
Hence, the envelope, $|\mathcal{R}(\tau)|$, of $\mathcal{R}(\tau)$ 
can be determined from
\begin{equation}
|\mathcal{R}(\tau)|\, = \, \sqrt {
R^2_{XX}(\tau) + R^2_{XY}(\tau) }.
\end{equation}

The total amount of ambiguity a waveform produces in the time domain is characterized by Woodward resolution constant, ${\scriptstyle {\Delta}} \tau$, defined by [17], [37]
\begin{eqnarray}
{\scriptstyle {\Delta}} \tau & \triangleq & 
\frac {\int _{-\infty} ^\infty \!  |{\cal R}(\tau)|^2\, d\tau}
{ |{\cal R}(0)|^2 }\\
&=& \frac {\int _0 ^\infty \!  S^2_X(\omega)\, d\omega}
{\frac {1}{2\pi} \mspace{-2mu}
\left [\int _0 ^\infty \! S_X(\omega)\, 
d\omega \right]^{\mspace{-1mu}2}  }\,. \nonumber
\end{eqnarray}
The constant ${\scriptstyle {\Delta}} \tau$ is expressed in units of time, and its reciprocal, $1/{\scriptstyle {\Delta}} \tau$, can be interpreted as the bandwidth (in Hertz) {\emph {occupied}} by the waveform{\footnote 
	{This definition of bandwidth is implicitly used in (40), when determining the number of degrees of freedom of a real waveform.}}.

In crosslation-based signal processing, complex autocorrelation, $\mathcal{R}(\tau)$, is replaced by complex crosslation 
${\cal A}(\tau)$. As a consequence, the corresponding Woodward resolution constant, ${\scriptstyle {\Delta}} \tau_{\cal C}$,  
assumes the form
\begin{eqnarray}
{\scriptstyle {\Delta}}\tau_{\cal C} & \triangleq & 
\frac {\int _{-\infty} ^\infty \!  |{\cal A}(\tau)|^2\, d\tau}
{ |{\cal A}(0)|^2 }\\
&=&
\frac {\int _0 ^\infty \! \omega^2 S^2_X(\omega)\, d\omega}
{\frac {1}{2\pi} \mspace{-2mu}
	\left [\int _0 ^\infty \!\omega\mspace{1mu} S_X(\omega)\, 
	d\omega \right]^{\mspace{-1mu}2}}\, . \nonumber
\end{eqnarray}
Owing to virtual high-frequency emphasis, the resolution constant 
${\scriptstyle {\Delta}} \tau_{\cal C}$ may be much smaller than 
 ${\scriptstyle {\Delta}} \tau$, especially in a case of power spectra 
 $S_X(\omega)$ with 'heavy' tails.

A useful measure of resolution gain, 
$C_{{\scriptscriptstyle {\Delta}} \tau}$, associated with crosslation-based processing, can be defined as follows
\begin{equation}
C_{{\scriptscriptstyle {\Delta}} \tau} \,\triangleq\,
\frac {{\scriptstyle {\Delta}}\tau}
{{\scriptstyle {\Delta}}\tau_{\cal C}} \,.
\end{equation}
For example, in the case of a Gaussian power spectral density, 
$\exp[-\omega^2/(2B^2)]$, 
$C_{{\scriptscriptstyle {\Delta}} \tau}=4/\pi$, whereas for a Laplacian density, $\exp(-|\omega|/W)$, the resolution gain
 $C_{{\scriptscriptstyle {\Delta}} \tau}$ equals $2$. 

In the following, resolution gains associated with crosslation-based processing will be determined for two classes of power spectrum density with its tail decaying at a rate that depends on a specified shape parameter.

\subsection{Butterworth Power Spectrum Density}
The Butterworth family of power spectrum densities is represented by the density
\begin{equation}
S_X(\omega) \,=\, \frac {1}
{1+ \left ( \frac {\omega}{W}  \right )^{\mspace{-2mu}2\kappa}}; 
\quad \kappa \geq 1, \,\, W >0
\end{equation}
where $\kappa\geq 1$ is a shape parameter, and $W$ is a positive constant related to the bandwidth. When $\kappa = 1$, the Butterworth density (53) has a Lorentzian (Cauchy) shape with a 'heavy' tail. As the value of $\kappa$  is increasing to infinity, the density (53) tends to a rectangular distribution, uniform in the interval $|\omega| < W$. 

Fig.\,13 shows the shape of Butterworth power spectrum density for four values of the shape parameter $\kappa$. As seen, the densities are decaying to zero at different rates, and the respective distributions exhibit 'light' as well as 'heavy' tails. 

\begin{figure}[]
     \centering
	\includegraphics[width=8.9cm]{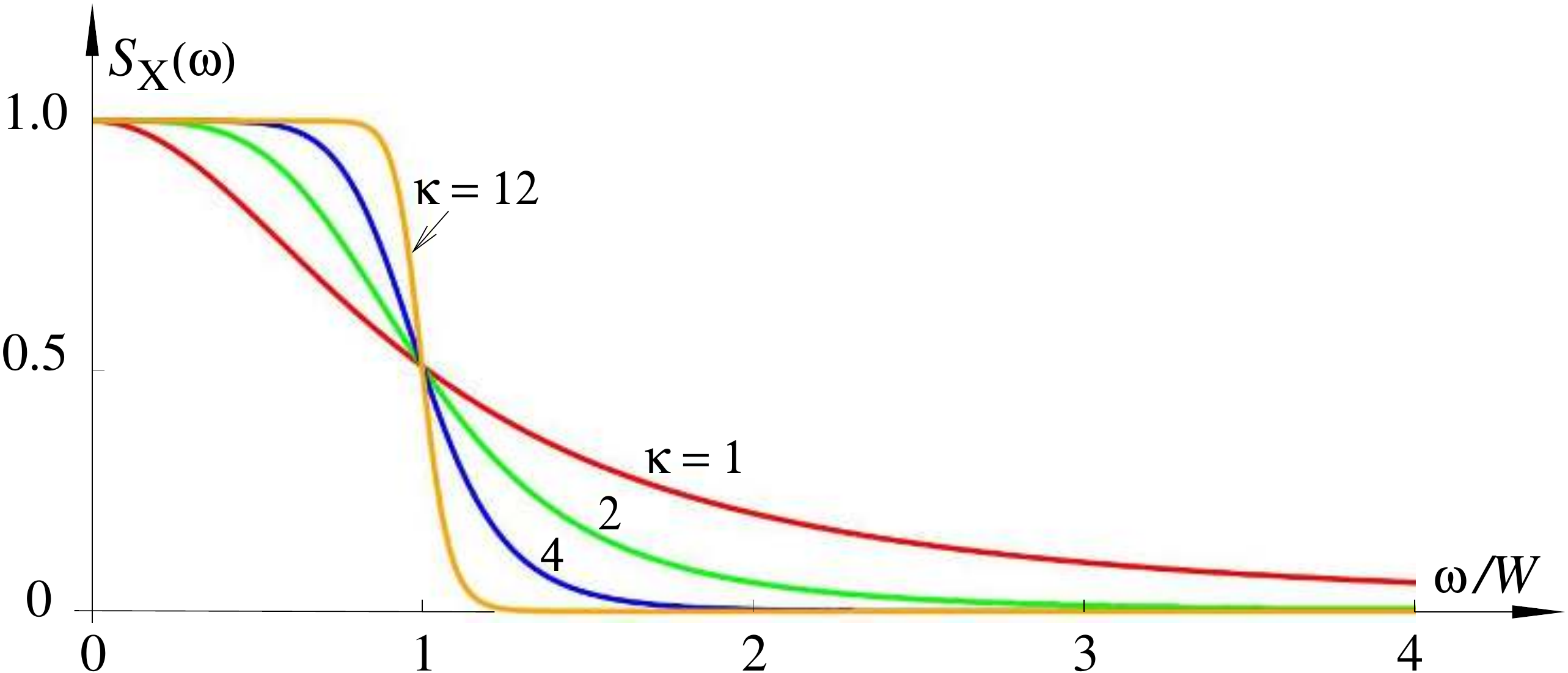}
	\vspace{-0.3cm}
	\caption{Dependence of Butterworth power spectrum density on shape parameter $\kappa$.}
\end{figure}

\begin{figure}[]
	\centering
	\includegraphics[width=7.6cm]{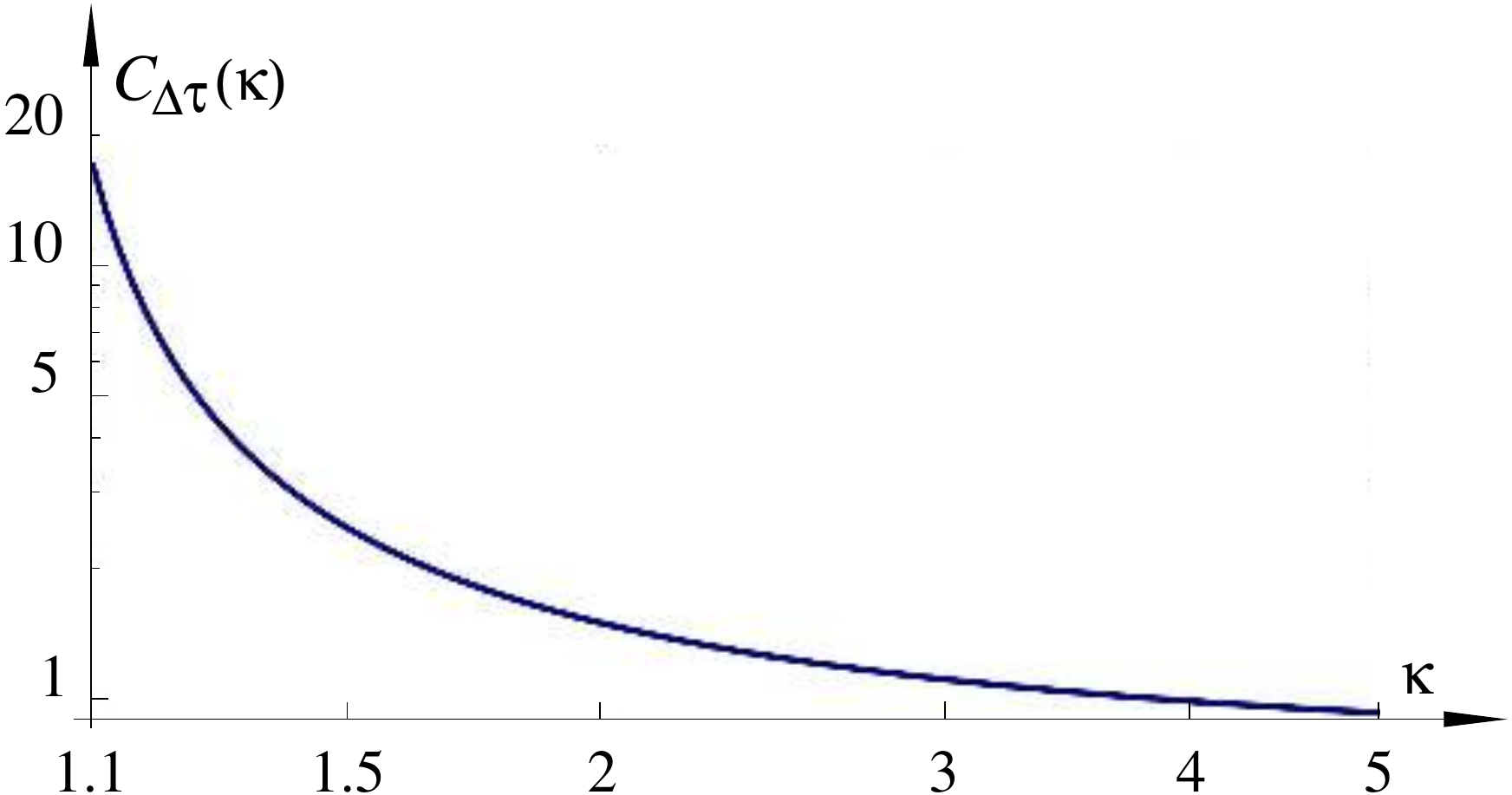}
	\vspace{-0.3cm}
	\caption{Resolution gain of crosslation-based processing determined for a Butterworth power spectrum density with shape parameter $\kappa$.}
\end{figure}

It can be shown that the resolution gain is given by
\begin{equation}
C_{{\scriptscriptstyle {\Delta}} \tau} (\kappa)\,= \,
 \frac {   {\mathsf{B}}(\frac {1} {2\kappa},\frac {4\kappa-1}{2\kappa})\,
 {\mathsf{B}}^2(\frac {1} {\kappa},\frac {\kappa-1}{\kappa})	              }
{  {\mathsf{B}}^2(\frac {1} {2\kappa},\frac {2\kappa-1}{2\kappa})\,
	{\mathsf{B}}(\frac {3} {2\kappa},\frac {4\kappa-3}{2\kappa})   }
\end{equation}
 where ${\mathsf{B}}(\cdot,\mspace{-3mu}\cdot)$ is the beta function;  ${\mathsf{B}}(\xi,\eta)\mspace{-1mu} =\mspace{-1mu} \Gamma(\xi)\Gamma(\eta)/\Gamma(\xi\mspace{-1mu}+\mspace{-1mu}\eta)$, 
 and $\Gamma(\cdot)$ is the gamma function [40, pp. 135--147].

The resolution gain, 
$C_{{\scriptscriptstyle {\Delta}} \tau}(\kappa)$, being less than unity, when $\kappa > 4$, rises rather slowly to 
$C_{{\scriptscriptstyle {\Delta}} \tau}(1.5) = 4\pi/(3\sqrt{3}) \approx 2.4$, and then, as $\kappa\! \rightarrow\! 1$, the gain is starting to increase rapidly to infinity, as shown in Fig.\,14. For example,   
$C_{{\scriptscriptstyle {\Delta}} \tau}(1.1) \approx 17$, and 
$C_{{\scriptscriptstyle {\Delta}} \tau}(1.05) \approx 53.5$. This 'super-resolution' is a manifestation of the phenomenon of virtual high-frequency emphasis.

\subsection{Band-Limited Power Spectrum Density}
Consider a Gaussian process $X(t)$ with a constant power spectral density $S_X(\omega)=\sigma^2_X\,\pi/W$, when $|\omega|<W$, and 
$S_X(\omega)=0$, elsewhere. Such a density is a special case of the Butterworth spectrum (53) when $\kappa$ tends to infinity.

\begin{figure}[]
	\centering
	\includegraphics[width=8.2cm]{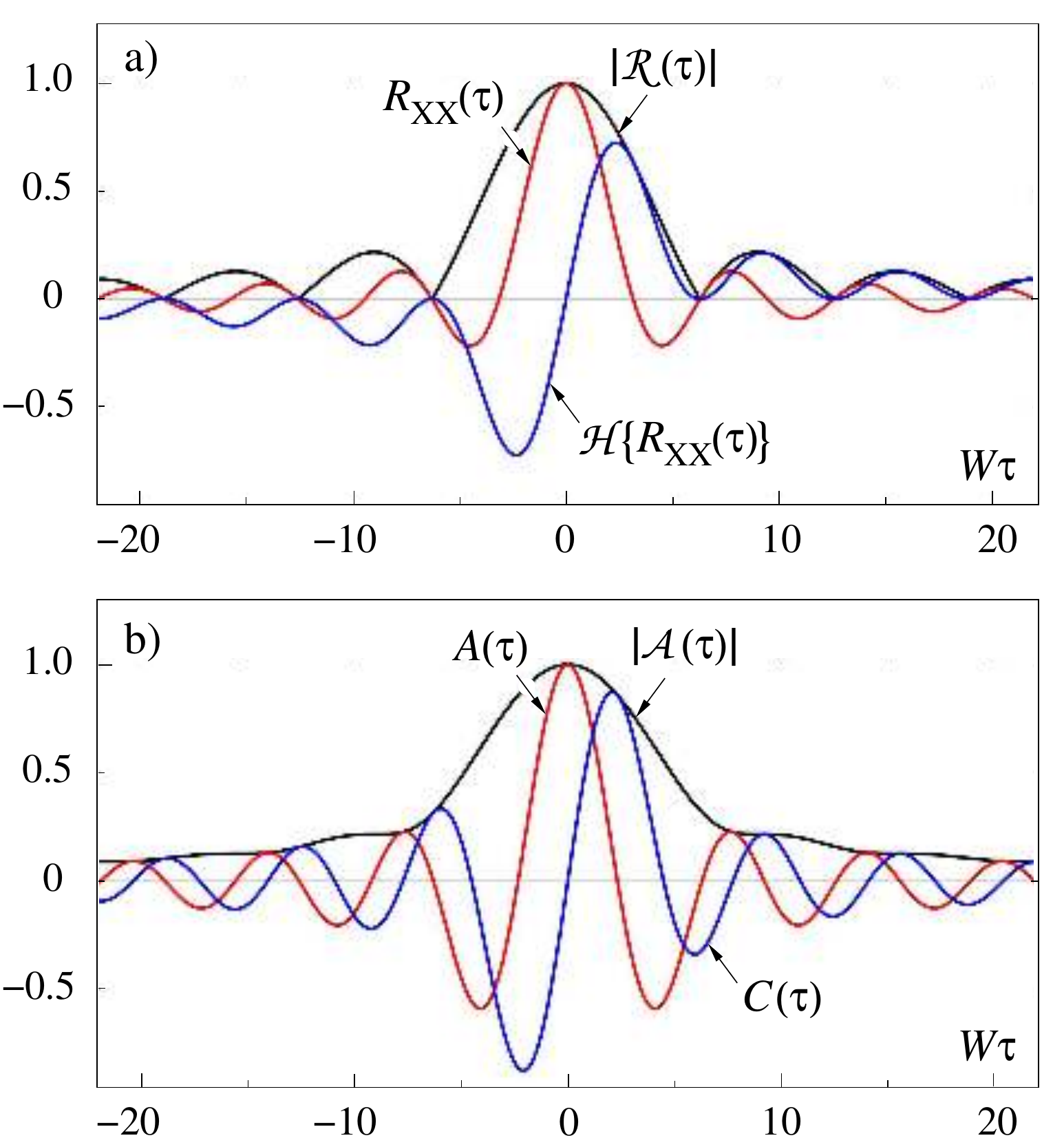}
	\vspace{-0.3cm}
	\caption{Envelope and components of complex correlation (a), and those of complex crosslation (b), determined for a Gaussian process 
		with a rectangular power spectrum density.}
\end{figure}

The autocorrelation function of the process is of the form
\begin{equation}
R_{XX}(\tau) \,=\, \sigma_X^2\mspace{1mu} \frac {\sin( W \tau )}{W\tau} \,.
\end{equation}
Since [31, eq. (7.114)]
\begin{equation}
R_{XY}(\tau) \,=\, \mathcal{H}\{R_{XX}(\tau)\} \, = \,
\sigma_X^2\mspace{1mu} \frac {1 - \cos( W \tau )}{W\tau}
\end{equation}
the envelope of the complex correlation function, $\mathcal{R}(\tau)$, is given by
\begin{equation}
|\mathcal{R}(\tau)|\mspace{1mu}  \,=\,\mspace{1mu} 
\sigma_X^2\mspace{-1mu} \left |
\frac {\sin( W \tau/2 )}{W\tau/2} \right | \, .
\end{equation}
The two components, $R_{XX}(\tau)$ and $R_{XY}(\tau)$, of the complex autocorrelation function $\mathcal{R}(\tau)$ and the resulting envelope $|\mathcal{R}(\tau)|$, when $\sigma_X^2 =1$, are shown in Fig.\,15\,a.

\begin{figure}[]
	\centering
	\includegraphics[width=7.7cm]{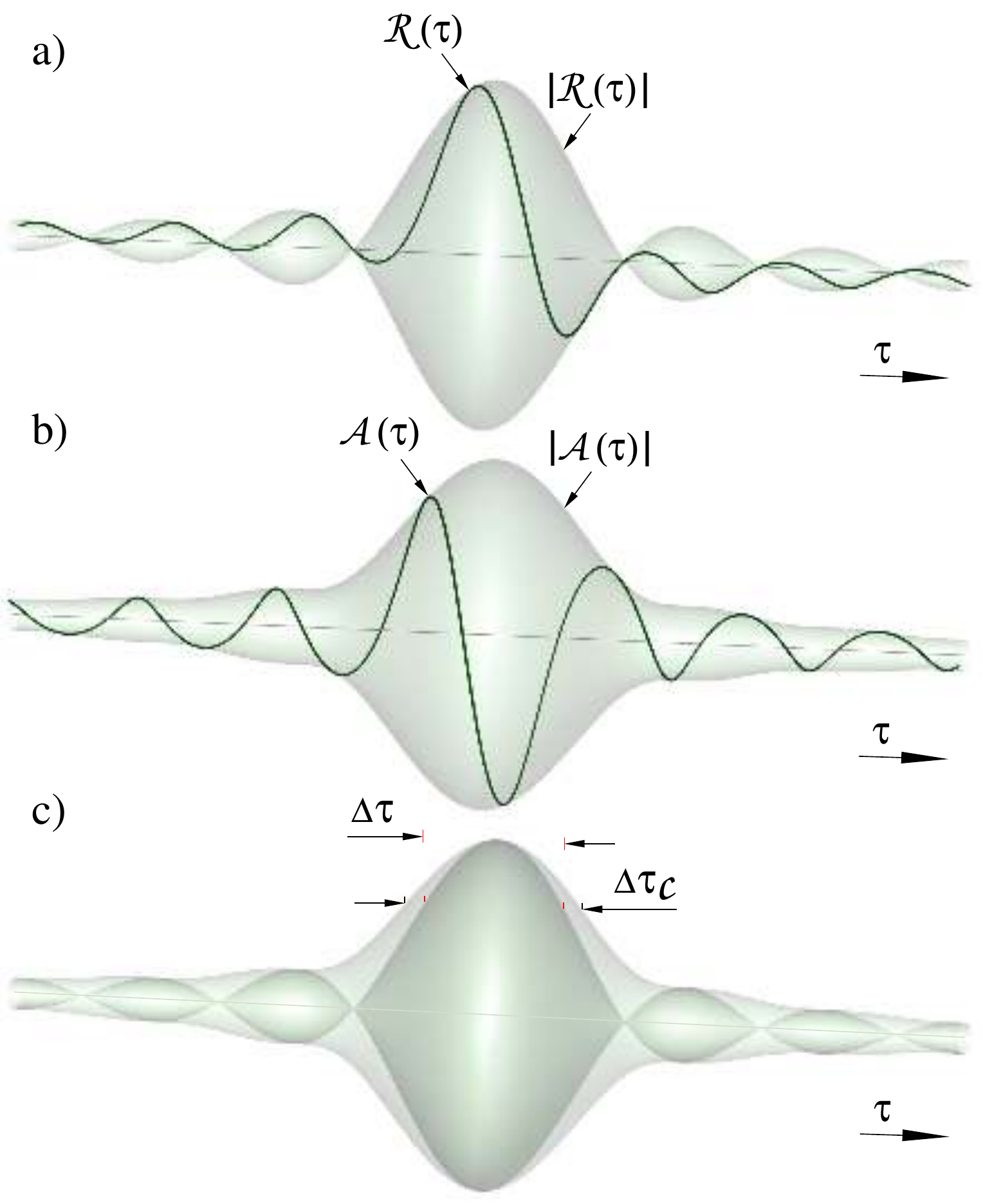}
	\vspace{-0.3cm}
	\caption{Correlation helix (a), crosslation helix (b), their superimposed spatial envelopes and respective resolution constants (c), determined for a Gaussian process 
		having a rectangular power spectrum density.}
\end{figure}

In the case of a band-limited process, $B_X = W/\sqrt{3}$, and the crosslation function, determined from (29) and (55), assumes the form 
\begin{equation}
C(\tau) \,=\, \sigma_X \sqrt{\frac {3 \pi}{2}} \mspace{1mu}
\frac {\mspace{1mu}\sin( W \tau ) - W\tau \cos( W \tau )}
{W^2\mspace{1mu}\tau^2}\, .
\end{equation}
The corresponding autoference function{\footnote{The order of differentiation and determining the Hilbert transform can be interchanged.}},
\begin{eqnarray}
A(\tau)&\!\!\! = \!\!\!& \mathcal{H}^{-1}\{C(\tau)\} \,=  \,
\sqrt{\frac {\pi}{2}} \frac {1}{B_X \sigma_X} \,
\mathcal{H}\{ R'_{XX}(\tau)\} \nonumber \\
&\!\!\!=\!\!\!& \sqrt{\frac {\pi}{2}} \frac {1}{B_X \sigma_X} \mspace{1mu}
\left[ \mathcal{H}\{ R_{XX}(\tau)\} \right]'
\end{eqnarray}
can be expressed as
\begin{equation} 
A(\tau) \, = \, \sigma_X \sqrt{\frac {3 \pi}{2}} \mspace{1mu}
\frac {\mspace{1mu}W\tau \sin( W \tau ) + \cos( W \tau ) -1}
{W^2\mspace{1mu}\tau^2} \, .
\end{equation}
Hence, the envelope of the complex crosslation function, 
$\mathcal{A}(\tau)$, is of the form
\begin{equation} 
|\mathcal{A}(\tau)| = \sigma_X \sqrt{\frac {3 \pi}{2}} \mspace{1mu}
\frac {\sqrt{W^2\mspace{1mu}\tau^2-2\mspace{1mu}W\tau \sin(W\tau) -2\cos(W\tau)+2}}
{W^2\mspace{1mu}\tau^2} .
\end{equation}
The two components, $A(\tau)$ and $C(\tau)$, of the complex crosslation function $\mathcal{A}(\tau)$ and the resulting envelope 
$|\mathcal{A}(\tau)|$ are shown in Fig.\,15\,b; the values of the components have been normalized so that $|\mathcal{A}(0)| = 1$. 

The time-resolution constants, ${\scriptstyle {\Delta}}\tau$ and 
${\scriptstyle {\Delta}}\tau_{\cal C}$, associated with correlation and crosslation are, respectively, given by
\begin{equation} 
{\scriptstyle {\Delta}}\tau   \,=\, 2 \pi/W, \quad
{\scriptstyle {\Delta}}\tau_{\cal C} \,=\,  8 \pi/(3W) .
\end{equation}
Therefore, in the case of a band-limited spectrum, 
$C_{{\scriptscriptstyle {\Delta}} \tau} = 3/4$, 
and crosslation-based processing will result in some resolution loss with respect to conventional correlation-based processing.

The correlation helix $\mathcal{R}(\tau)$, crosslation helix $\mathcal{A}(\tau)$, and their envelopes, each represented by a surface of revolution, are shown, respectively, in Fig.\,16\,a and Fig.\,16\,b. The two superimposed envelopes, $|\mathcal{R}(\tau)|$ and $|\mathcal{A}(\tau)|$, are shown in Fig.\,16\,c; for comparison purposes, the respective peak values have been adjusted so that $|\mathcal{A}(0)|=|\mathcal{R}(0)|$. The correlation envelope exhibits an oscillatory character, whereas the crosslation envelope is broader and almost unimodal. As seen, the adjusted crosslation envelope may be regarded as a kind of 'envelope' of the correlation envelope.
	
\subsection{Lorentzian Power Spectrum Density}
When the shape parameter $\kappa=1$, the Butterworth power spectrum (53) becomes a Lorentzian spectrum with density
\begin{equation}
S_L(\omega) \,=\, \frac {2\mspace{1mu} \sigma^2_X}
{W \!\left[1+ \left( \frac {\omega}{W} \right )^{\!2} \right] 
}; \quad W > 0.
\end{equation}
 The corresponding autocorrelation function has a Laplacian shape, $\sigma^2_X  \exp(-W |\tau|)$, where $\sigma^2_X$ is the variance. 

The Lorentzian spectrum plays a prominent role in the theory of random   frequency modulation. While for slow modulation the spectral density of the modulated signal follows the probability distribution of modulating noise, fast (wideband) modulation results in a Lorentzian spectrum of diffusion [38].

The Lorentzian spectrum can also be used as a representative model of power spectra of various physical phenomena such as galactic noise, ambient acoustic or seismic noise etc.

\subsubsection*{Modified Lorentzian Spectrum}
In practical systems, the Lorentzian spectrum of an intercepted waveform  will always be limited by transfer functions of various components of the system. Therefore, for the purpose of analysis,  a modified power spectrum density 
\begin{equation}
S_X(\omega) \,=\, \frac {2\mspace{1mu} \sigma^2_X (1+\gamma)}
{W\mspace{-1mu}\gamma\mspace{1mu}\Big[1+ \big( \frac {\omega}{W} \big )^{\!2} \Big] 
\Big[1+ \big ( \frac {\omega}{W\gamma} \big )^{\!2} \Big]
}
\end{equation}
will be used, where the parameter $\gamma$ is a measure of bandwidth limitation. The underlying Lorentzian spectrum will remain undistorted when $\gamma \rightarrow  \infty$.

The autocorrelation function $R_{XX}(\tau)$, corresponding to the spectrum (64), assumes the form{\footnote {The expressions (65) and (66) have been derived with the use of {\tt {Maxima}}, a computer algebra system.}}
\begin{equation}
R_{XX}(\tau) \,=\, \frac {\sigma^2_X  
\big[
\gamma\mspace{1mu} \exp(-W|\tau|)-
\exp(-\mspace{1mu}W\mspace{-1mu}\gamma\mspace{1mu}|\tau|)
\big]}
{\gamma-1}\, .
\end{equation}

\begin{figure}[]
	\centering
	\includegraphics[width=8.cm]{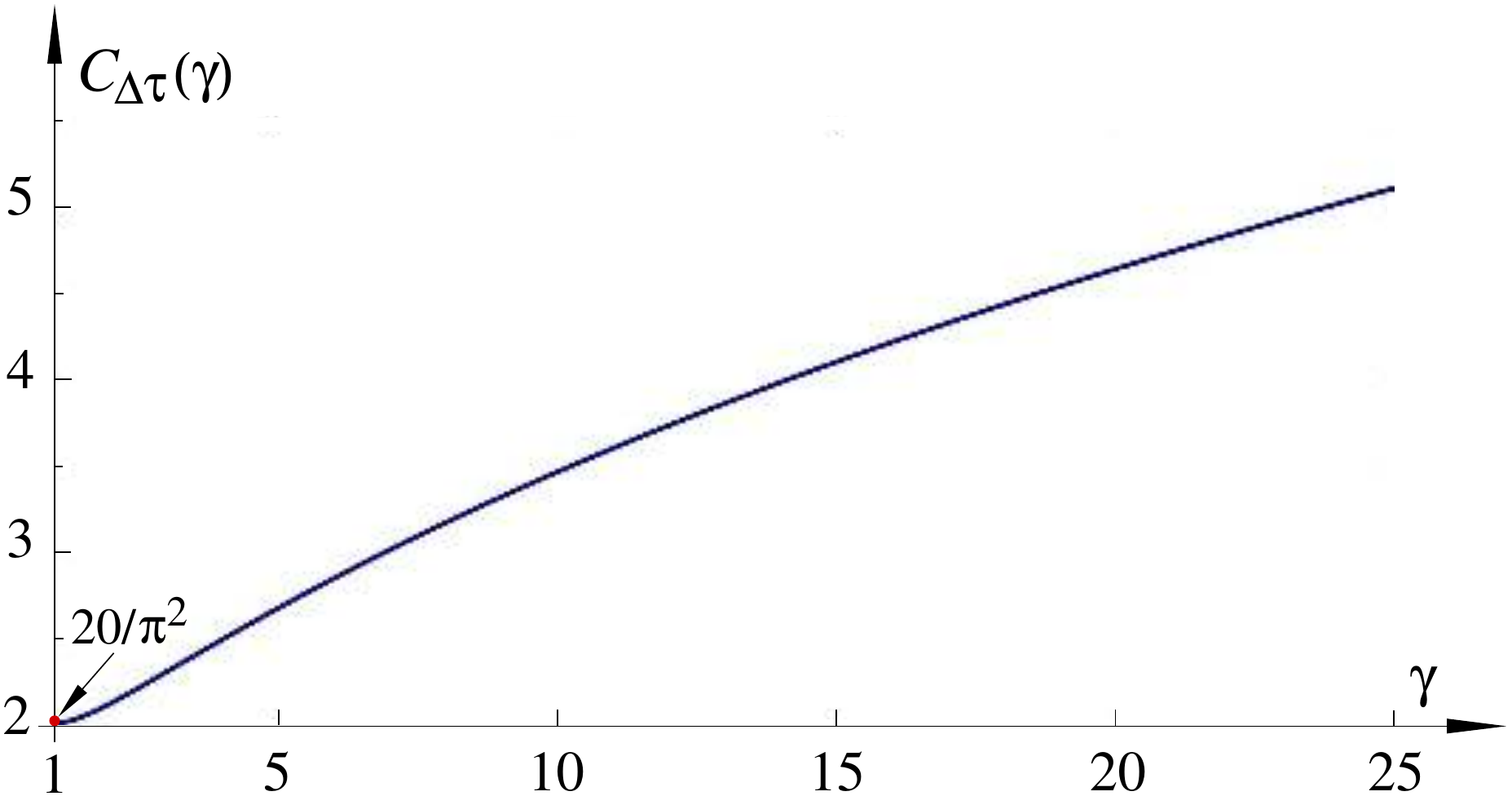}
	\vspace{-0.3cm}
	\caption{Resolution gain of crosslation-based processing determined for 
		a process with a modified Lorentzian spectrum.}
\end{figure}

It can be shown that the resolution gain, associated with crosslation-based processing, can be determined from
\begin{equation}
C_{{\scriptscriptstyle {\Delta}} \tau} (\gamma)\,= \,
\frac {4\mspace{1mu}(\gamma^2+3\gamma+1)\mspace{-1mu} \ln^2\gamma}
{\pi^2(\gamma-1)^2}\, .
\end{equation}
Fig.\,17 shows the crosslation gain $C_{{\scriptscriptstyle {\Delta}} \tau} (\gamma)$ as a function of the bandwidth limiting parameter $\gamma$. As expected, when $\gamma \rightarrow \infty$, the gain tends to infinity. However, in the case of significant bandwidth reduction, when $\gamma \rightarrow 1$, $C_{{\scriptscriptstyle {\Delta}} \tau} (\gamma) \rightarrow 20/\pi^2 \approx 2.03$.

The rms bandwith $B_X$, defined by (28), of a Lorentzian spectrum (63) is infinite; hence, the zero-crossing rate, $B_X/\pi$ of Gaussian noise with spectrum (63) is also, in theory, infinite. Since in the case of a Lorentzian spectrum, the number of degrees of freedom $\Lambda$, given by (40), is finite, the value $\gamma^*$ of band limiting parameter $\gamma$ can be so selected that $\Lambda(\gamma^*)$ will be equal to the number of zero crossings $n_c$, observed in the interval $T$.

Fig.\,18 shows the plots of $\Lambda/(WT)$ and $n_c/(WT)$ in the interval, $1 < \gamma <10$, split into two regions by the value of $\gamma^* \approx 5.56$. Accordingly, when $\gamma < \gamma^*$, the noise waveform is underdetermined by its zero crossings, whereas in the region $\gamma > \gamma^*$, the waveform is overdetermined.

\begin{figure}[]
	\centering
	\includegraphics[width=8.cm]{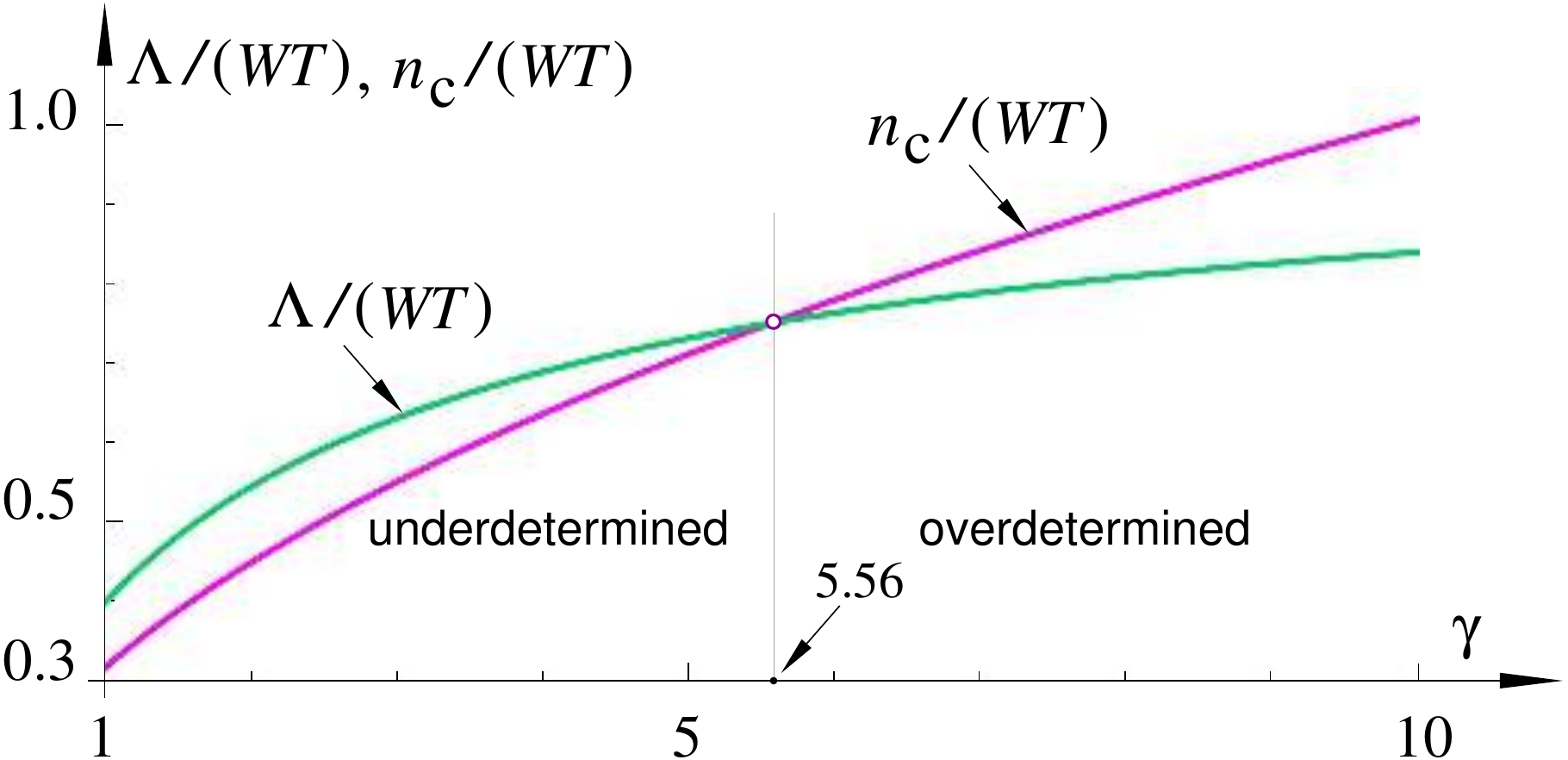}
	\vspace{-0.3cm}
	\caption{Regions in which a Gaussian process with a modified Lorentzian spectrum is underdetermined and overdetermined by its zero crossings.
		}
\end{figure}

The orthogonal component $\mathcal{H}\{R_{XX}(\tau)\}$ of a complex correlation function $\mathcal{R}(\tau)$ can be determined from (65) by exploiting the relationship [39]
\begin{equation*}
\mathcal{H}\{\mathrm{e}^{-\gamma|\tau|}\}  =
\frac {\mathrm{sgn}(\tau)}{\pi}\mspace{-1mu} \left[
 \mathrm{e}^{\gamma|\tau|}  E_1(\gamma|\tau|)    +               
 \mathrm{e}^{-\gamma|\tau|} \mathrm{Ei}(\gamma|\tau|)\right]
\end{equation*}
where $E_1(\tau)$ and ${\mathrm{Ei}}(\tau)$ are exponential integrals 
[40, pp. 149--157]. The two components, $R_{XX}(\tau)$ and $\mathcal{H}\{R_{XX}(\tau)\}$, of the complex autocorrelation function $\mathcal{R}(\tau)$ and the resulting envelope $|\mathcal{R}(\tau)|$, when $\sigma_X^2 =1$ and $\gamma=5$, are shown in Fig.\,19\,a.

\begin{figure}[]
	\centering
	\includegraphics[width=8.2cm]{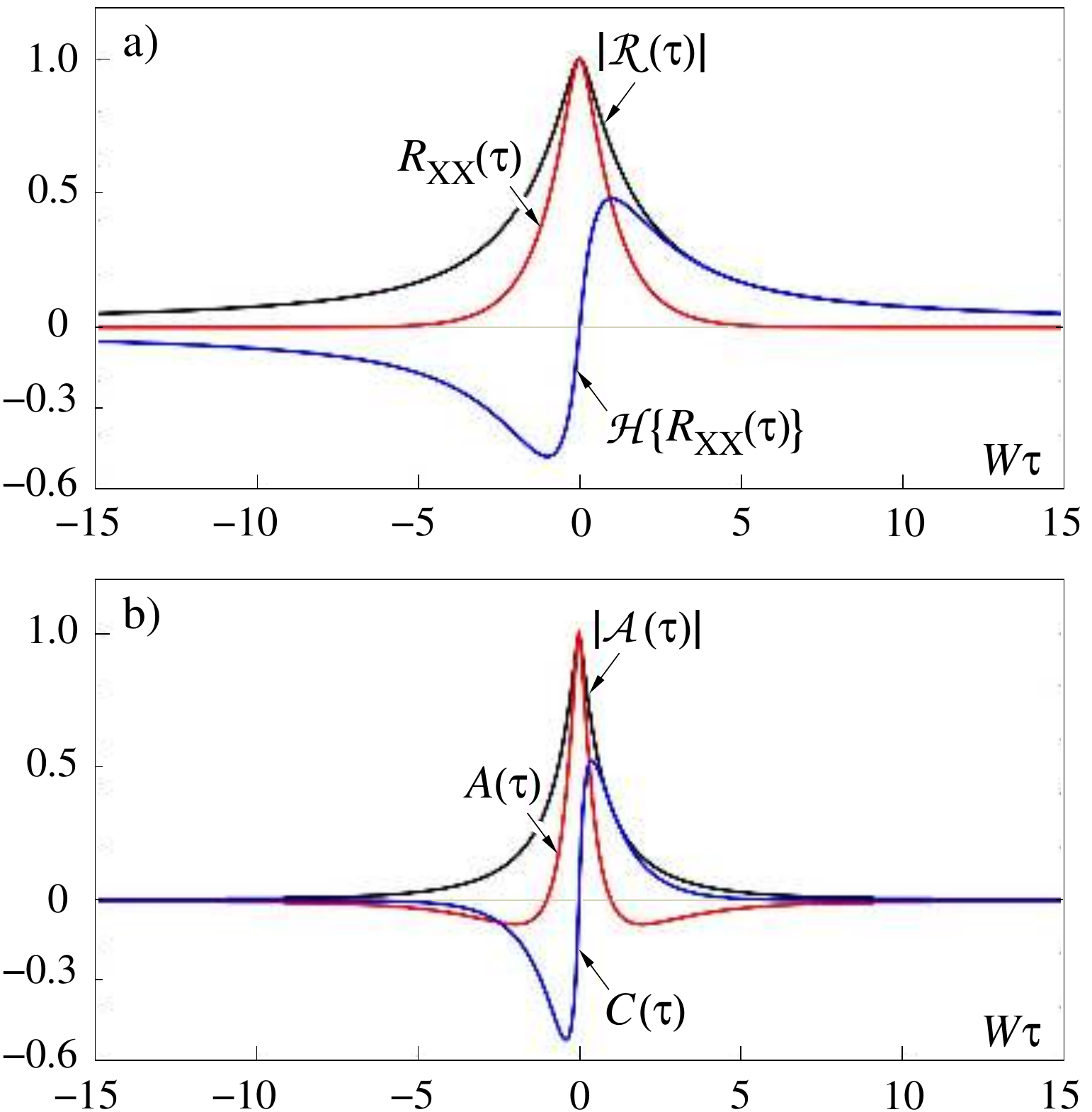}
	\vspace{-0.3cm}
	\caption{Envelope and components of complex correlation (a), and those of complex crosslation (b), of a process 
		with spectrum (64), when $\gamma=5$.}
\end{figure}

The relationship [39]
\begin{equation*}
\mathcal{H}\{{\mathrm{sgn}}(\tau)\, \mathrm{e}^{-\gamma|\tau|}\} =
\frac {1}{\pi}\mspace{-1mu} \left[
\mathrm{e}^{-\gamma|\tau|} \mathrm{Ei}(\gamma|\tau|) -
\mathrm{e}^{\gamma|\tau|}  E_1(\gamma|\tau|)             
\right]
\end{equation*}
has been used in conjunction with (29) to determine the two components, $C(\tau)$ and $A(\tau)$, of the complex crosslation function $\mathcal{A}(\tau)$. The components and the envelope $|\mathcal{A}(\tau)|$, for $\gamma=5$, are shown in Fig\, 19\,b; the values of the components have been normalized so that $|\mathcal{A}(0)| = 1$.

\begin{figure}[]
	\centering
	\includegraphics[width=8.2cm]{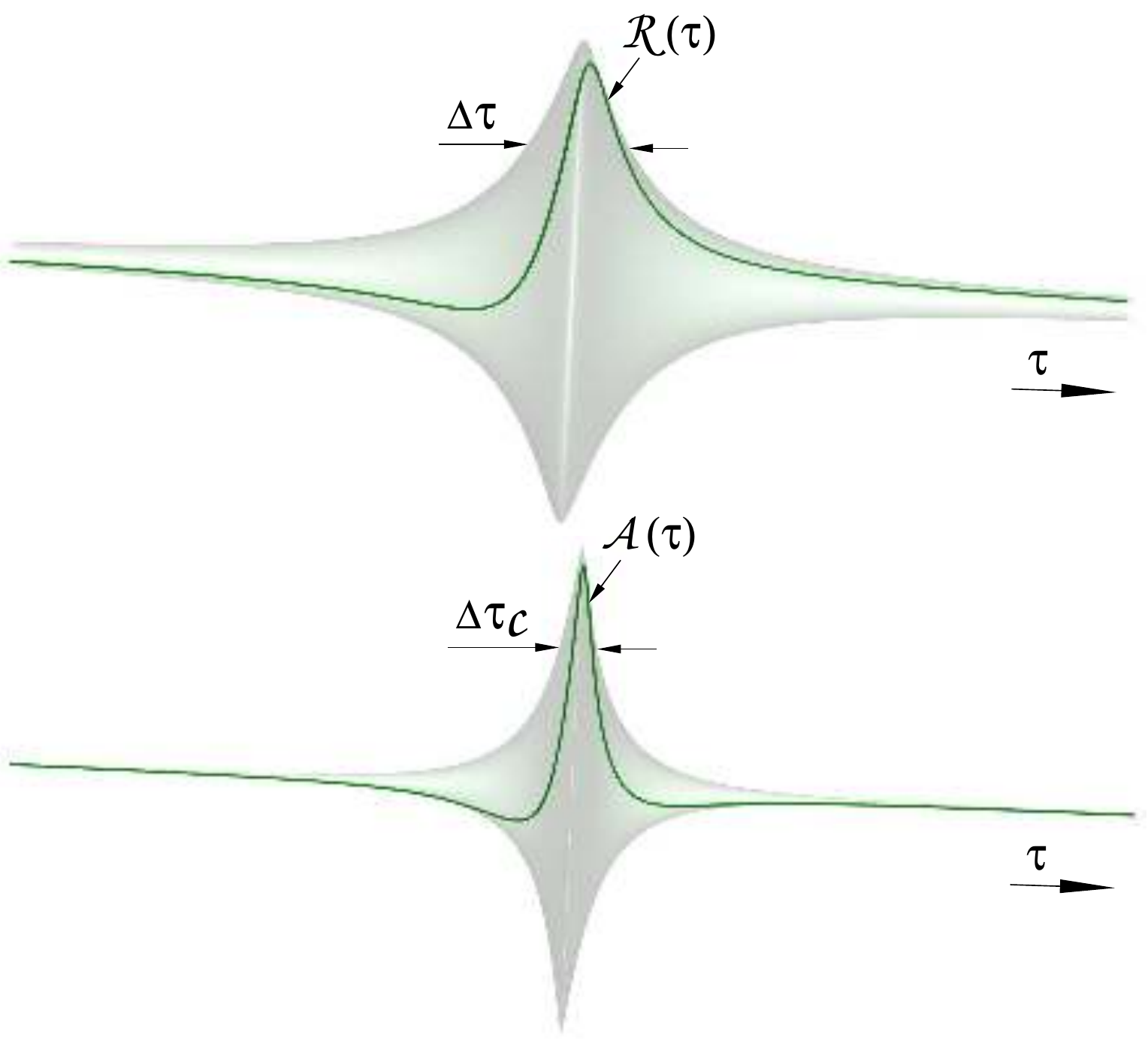}
	\vspace{-0.3cm}
	\caption{Correlation helix ${\cal R}(\tau)$, crosslation helix ${\cal A}(\tau)$, their respective envelopes and resolution constants, determined for a process with a modified Lorentzian spectrum ($\gamma=5$).}
\end{figure}

The correlation helix $\mathcal{R}(\tau)$, crosslation helix $\mathcal{A}(\tau)$, and their envelopes, each represented by a surface of revolution, are shown in Fig.\,20. In the considered case, $\gamma\!=\!5$, and the corresponding resolution gain $C_{{\scriptscriptstyle {\Delta}} \tau} (\gamma)$ of crosslation-based processing is approximately equal to $2.7$.

\subsection{The Cram\'er-Rao Bound}
The resolution constant (50) characterises a waveform in a noiseless case; however, in practice, a time-delayed replica of a reference waveform $x(t)$ is always corrupted by external noise. When the noise is relatively weak, the  Cram\'er-Rao (CR) bound on the variance of any unbiased estimator $\widehat{\small{\Delta}}$ of unknown time delay ${\small{\Delta}}$ can be expressed as [41]
\begin{equation} 
{\mathrm {var}}\{\widehat{\small{\Delta}}      \} \,\, \geq
\, \,\frac{\sigma_n^2}
{\sum_{m=1}^{m_s} \mspace{-2mu}
\Big[\!\! \left . \frac {\partial \mspace{1mu}x(t)}{\partial \mspace{1mu}t} \right |_{t=t_m} \!
 \Big ]^{\mspace{-1mu}2} }
\end{equation}
where $\sigma_n^2$ is the noise variance, and 
$\{t_m;\,\, m=1, \ldots, m_s \}$ are equidistant sampling times of a low-pass waveform $x(t)$. 

When the distance between consecutive samples is assumed to be infinitesimally small, the sum in (67) is approximated by an integral [41], and 
\begin{equation} 
{\mathrm {var}}\{\widehat{\small{\Delta}}      \} \,\, \geq \,
\, \frac{\sigma_n^2}
{ \frac {m_s}{T}\! \int_0^T  \!
\left	[  \frac {\partial\mspace{1mu} x(t)}{\partial\mspace{1mu} t}  
\right	]^{\mspace{-2mu}2} \mspace{-4mu} dt}
\end{equation}
where $T$ is an observation interval. Since a low-pass waveform $x(t)$ is a realization of a stationary and ergodic process $X(t)$,
\begin{eqnarray} 
\lim_{T \rightarrow \infty} \frac {1}{T}\!
\int_0^T \! 
\left[\frac {\partial\mspace{1mu} x(t)}{\partial\mspace{1mu} t}\right]^{\!2} \mspace{-5mu}dt
&\!\!\!=\!\!\!& R_{X'X'} (0)  \nonumber \\
&\!\!\!=\!\!\!& -R''_{XX}(0) \,= \, \sigma_X^2 \mspace{1mu}  B^2_X .\qquad
\end{eqnarray}
As a consequence, the CR bound (68) reduces to
\begin{equation} 
{\mathrm {var}}\{\widehat{\small{\Delta}}      \} \,\, \geq \,\,
  \frac{\sigma_n^2} {\Lambda\mspace{1mu} \sigma_X^2 \mspace{1mu}  B^2_X } 
\end{equation}
where it is assumed that the number of samples $m_s$ is {\em {equivalent}} to the number of degrees of freedom $\Lambda$ (i.e. the number of uncorrelated samples) characterising the waveform.

In the case of crosslation-based signal processing, the bound (67) assumes the form
\begin{equation}  
{\mathrm {var}}\{   {\widehat{\small{\Delta}}}_{\mathcal C}      \} 
\,\, \geq\,
\, \frac{\sigma_n^2}
{\sum_{i=1}^{n_c}  \mspace{-2mu}
	\Big[\!\! \left . \frac {\partial\mspace{1mu} x(t)}{\partial\mspace{1mu} t} \right |_{t=t_i} \!
	\Big ]^{\mspace{-1mu}2} }
\end{equation}
where  $\{t_i;\,\, i=1, \ldots, n_c \}$ are zero-crossing times occurring in a time interval $T$. 
From the Slepian model (30), it follows that the absolute slope, 
$|{\partial x(t)}/{\partial t}|$, at a zero crossing is a realization of a Rayleigh rv $U$ with the second moment equal to 
$2\mspace{1mu}\sigma_X^2 \mspace{1mu} B^2_X $. 
 Hence, when a waveform is underdetermined by its zero crossings,
\begin{equation} 
{\mathrm {var}}\{   {\widehat{\small{\Delta}}}_{\mathcal C}      \} \,\, \geq
\,\, \frac{\sigma_n^2} {2\mspace{1mu} n_c\mspace{1mu} \sigma_X^2 \mspace{1mu}  B^2_X },
 \qquad n_c < \Lambda 
\end{equation} 
whereas, when a waveform is determined or overdetermined by its zero crossings, 
\begin{equation} 
{\mathrm {var}}\{   {\widehat{\small{\Delta}}}_{\mathcal C}      \} \,\, \geq
\,\, \frac{\sigma_n^2} {2\mspace{1mu} \Lambda \mspace{1mu}\sigma_X^2  B^2_X }\, = \,
\frac {1}{2}\, {\mathrm {var}}\{\widehat{\small{\Delta}}      \},
\quad n_c \ge \Lambda .
\end{equation}
Therefore, in the case of crosslation-based processing of a random waveform, the CR bound (73) is equal to a half of that, (70), associated with correlation-based processing{\footnote {An analogous effect can be observed in the case of a sinusoidal waveform.}}. 

The slope at a zero crossing,
\begin{equation}
{\vartheta}_i \,\mspace{1mu} \triangleq \, 
\left . \frac {\partial \mspace{1mu} x(t)}{\partial \mspace{1mu} t} \right|_{t=t_i} 
\end{equation}
referred to as a {\em {local slew rate}}, is a measure of the speed with which a waveform $x(t)$ can change its values while crossing a zero level. Observables $\{|{\vartheta}_i|\}$ are realizations of a Rayleigh rv $U$ with the mean equal to the {\emph {global}} slew rate, $\bar{\vartheta} = \sqrt{\pi/2}\mspace{2mu}B_X \sigma_X \equiv \mathrm{E}\{U\}$.  

The phenomenon of local slew rate has been demonstrated experimentally. Fig.\,21 is a long-exposure photograph of an analogue oscilloscope screen showing crossjectories $\{x_k^+(\tau)\}$ of a reference random waveform (top panel) and trajectories $\{x_k^+(\tau-\small{\Delta})\}$ of a delayed replica  of the same waveform, observed at the output of a delay line (bottom panel). The oscilloscope was triggered by zero upcrossings of the reference waveform. As seen, the slopes of crossjectories at $\tau = 0$, and also those of corresponding trajectories at $\tau =\small{\Delta}$, are spread around their respective mean values $\bar{\vartheta}$.

\begin{figure}[]
	\centering
	\includegraphics[width=7cm]{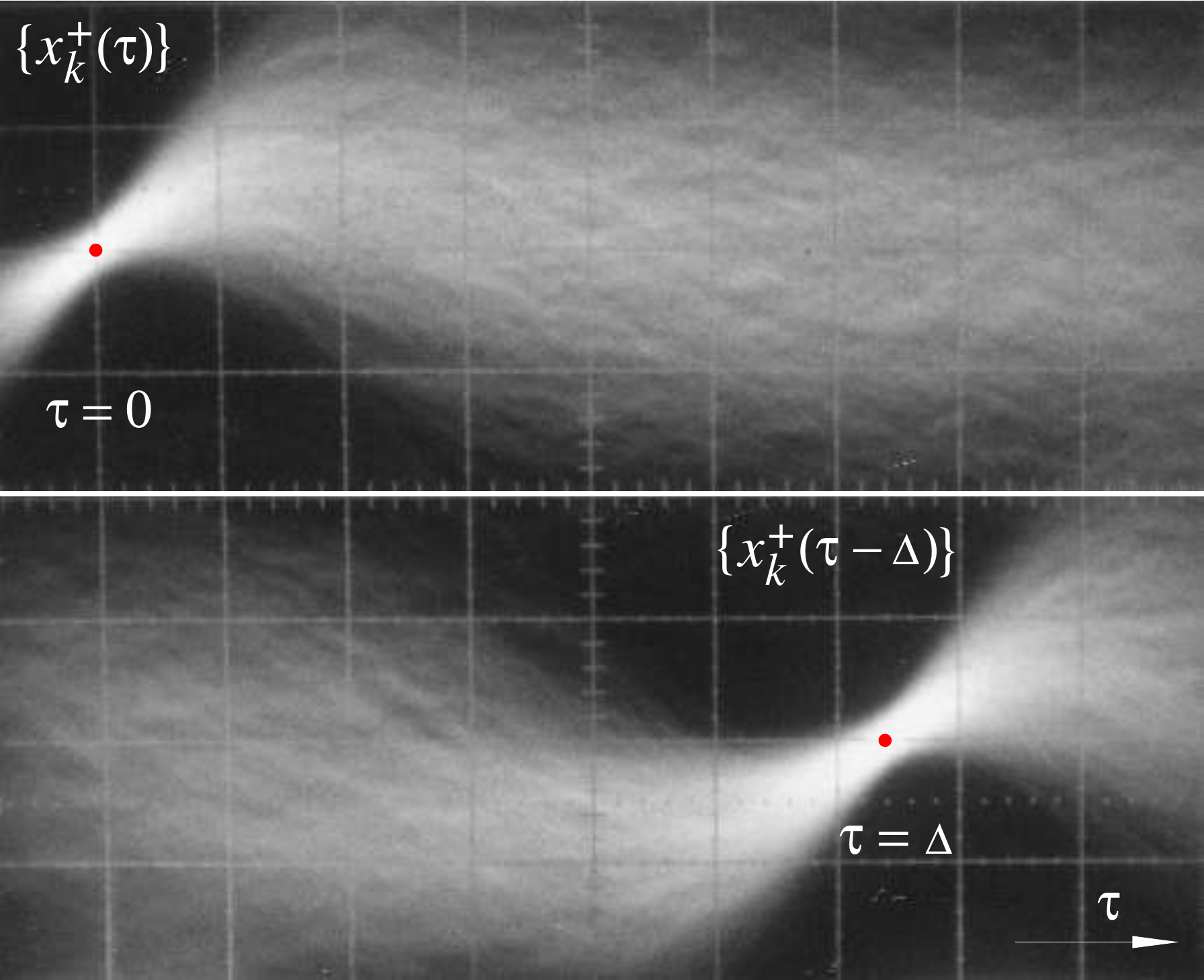}
	\vspace{-0.0cm}
	\caption{Crossjectories $\{x_k^+(\tau)\}$ associated with zero upcrossings (top) and their delayed replicas, $\{x_k^+(\tau-\small{\Delta})\}$, (bottom).}
\end{figure}

Information about local slew rate can be exploited to modify empirical crosslation (11) and empirical autoference (37) functions as follows
\begin{eqnarray}
\widehat{C}_{\vartheta}(\tau) &\!\! \triangleq \!\! &
\frac {1}{n_c} \sum_{i=1}^{n_c}
{\vartheta}_i \mspace{2mu} 
x(t_i+\tau) \\
\widehat{A}_{\vartheta}(\tau) &\!\! \triangleq \!\! &
\frac {1}{n_c} \sum_{i=1}^{n_c}
{\vartheta}_i \mspace{2mu} 
y(t_i+\tau) 
\end{eqnarray}
where ${\vartheta}_i$ is a local slope determined from the reference waveform in accordance with (74).

For example, the operation of {\em {slew-rate matching}} in (75) can be performed by a crosslator of Fig.\,5, modified in such a way that the  buffer/inverters {\sf {B}} are replaced by amplifiers whose gain is controlled by a value proportional to a {\emph {signed}} slope ${\vartheta}_i$. The required value of ${\vartheta}_i$ is provided, at output $\psi_i$, by a suitably modified zero-crossing detector.

When a waveform is overdetermined by its zero crossings, i.e. when $n_c > \Lambda$, the CR bound (73) can be further reduced by retaining crossings associated with $\Lambda$ largest absolute slope values $|{\vartheta}_i|$, and discarding $(n_c\mspace{-2mu}-\mspace{-2mu}\Lambda)$ remaining crossings.
 
Such a trimming procedure can be replaced by a more practical decimation scheme in which retained crossings are associated with absolute slope values  $|{\vartheta}_i|$ that have exceeded a predetermined threshold $\eta_{\Lambda}$. The value of $\eta_{\Lambda}$ is so selected as to retain, {\em {on average}}, $\Lambda$ crossings. In this case, the CR bound reduction will result from the fact that the second moment of a truncated Rayleigh rv $U_{\Lambda}$, with realizations restricted to the interval $(\eta_{\Lambda}, \infty )$, is greater than $2\mspace{1mu}\sigma_X^2  B^2_X$. 

The above decimation scheme can also be exploited when a waveform is undetermined by its zero crossings, i.e. when $n_c < \Lambda$. In either case, the proposed decimation will improve the performance of slew-rate matching in (75) and (76).

\section{Conclusions}
The techniques presented in this paper are particularly well suited to dealing with problems in which a reference signal waveform can be uniquely identified and its zero crossings reliably extracted to control the sampling pattern of a waveform being processed. The are three main classes of such problems:\\
1. waveform analysis -- a reference waveform is also the waveform being processed to determine its power spectrum, components of the crosslation helix or the Nyquist plot;\\
2. active systems -- a reference waveform is the waveform generated by the system for illumination (interrogation) of a sensed environment, and a waveform being processed is the response of the environment;\\
3. passive systems exploiting 'day-light' illumination [5] -- a reference illuminating waveform (such as galactic noise) is captured by one sensor whereas a response waveform to be processed is supplied by a separate sensor.

In passive sensing, when no distinct reference waveform is available, the presented techniques can be modified in such a way that each of the two captured waveforms is used as a reference to process the other waveform. Then, the two such obtained zero-crossing interferograms are suitably combined to produce a single resultant interferogram containing information about the environment being sensed [42]. 

\section*{Acknowledgement}
This paper is based, in part, on research work supported by Visual Information Laboratory (closed in 2010), Guildford, Mitsubishi Electric Europe B.V., U\,K.

The author gratefully acknowledges the many stimulating discussions and useful suggestions provided by Dr. Paul Ratliff, Prof. Miroslaw Bober, and  
Dr. Wojciech Machowski.

\end{document}